\documentclass[preprint]{aastex}
\usepackage{xspace}

\def\gcc{${\rm g\ \hbox{cm}^{-3}}$\xspace}

\def\photpersec{${\rm \hbox{photons}\ s^{-1}}$\xspace}

\def\kms{${\rm \hbox{km}\ s^{-1}}$\xspace}

\def\sun {${}_\odot$}
\def\Msun{M\sun\xspace}

\def\msun{\hbox{M}_\odot}
\def\mch{${\rm M_{ch}}$\xspace}
\def\Mch{\mch}
\def\gtsim{\,\hbox{\raise 0.2em\hbox{$ > $}\kern -0.8em \lower 0.7ex\hbox{$\sim$}}\,}
\def\ltsim{\,\hbox{\raise 0.2em\hbox{$ < $}\kern -0.8em \lower 0.7ex\hbox{$\sim$}}\,}

\def\nifsx{${}^{56}$Ni\xspace}
\def\cofsx{${}^{56}$Co\xspace}
\def\fefsx{${}^{56}$Fe\xspace}
\def\ka{K$\alpha$\xspace}
\newcommand\eqn[1]{(\ref{#1})}
\def\msunperyear{M_\odot-{\rm yr^{-1}}}
\def\cf{{\it c.f.}\xspace}
\def\eg{{\it e.g.}\xspace}

\begin{document}

\title{A Test for the Nature of the Type Ia Supernova Explosion Mechanism}

\shorttitle{Test for SN~Ia Explosion Mechanism}
\shortauthors{Pinto, Eastman, \& Rogers}

\author{Philip A. Pinto\altaffilmark{1}}
\affil{Steward Observatory\\
University of Arizona\\Tucson, AZ 85721 USA}
\email{ppinto@as.arizona.edu}

\author{Ronald G. Eastman\altaffilmark{2}}
\affil{Lawrence Livermore National Laboratory\\
Livermore, CA 94550 USA}
\email{reastman@llnl.gov}

\and

\author{Tamara Rogers\altaffilmark{1}}
\affil{Department of Astronomy and Astrophysics \\
University of California Santa Cruz, \\
Santa Cruz CA 95064 USA}
\email{tami@ucolick.org}

\altaffiltext{1}{Lawrence Livermore National 
Laboratory, Livermore CA 94550 USA}
\altaffiltext{2}{Department of Astronomy and Astrophysics, 
University of California, Santa Cruz, Santa Cruz CA 95064 USA}


\setcounter{footnote}{0}

\begin{abstract}
Currently popular models for Type Ia supernov\ae\ (SNe~Ia) fall into
two general classes. The first comprises explosions of nearly pure
carbon/oxygen (C/O) white dwarfs at the Chandrasekhar limit which
ignite near their centers. The second consists of lower-mass C/O cores
which are ignited by the detonation of an accreted surface helium
layer. Explosions of the latter type produce copious Fe, Co and Ni \ka
emission from \nifsx and \cofsx decay in the detonated surface layers,
emission which is much weaker from Chandrasekhar-mass models.  The
presence of this emission provides a simple and unambiguous
discriminant between these two models for SNe~Ia. Both mechanisms may
produce $0.1-0.6\msun$ of \nifsx, making them bright $\gamma$-ray line
emitters. The time to maximum brightness of \nifsx decay lines is
distinctly shorter in the $M<M_{ch}$ class of model ($\sim15$ days)
than in the \Mch model ($\sim30$ days), making $\gamma$-ray line
evolution another direct test of the explosion mechanism.  It should
just be possible to detect K-shell emission from a sub-\Mch explosion
from SNe~Ia as far away as the Virgo cluster with the XMM Observatory.
A 1 to 2 $(\hbox{meter})^2$ X-ray telescope such as the proposed Con-X
Observatory could observe K$\alpha$ emission from $M<M_{ch}$ SNe~Ia in
the Virgo cluster, providing not just a detection, but high-accuracy
flux and kinematic information.

\end{abstract}


\newpage

\section{Introduction}

The luminosity of Type Ia supernovae (SNe~Ia), the brightest stellar
explosions, arises in current successful models not from the energy of
the explosion itself but from the radioactive decay of \nifsx.
Following \citet{Arnett69}, in all current models these explosions are
the thermonuclear incineration of a white dwarf. The very compact
nature of this progenitor ensures that the explosion energy is
efficiently converted to kinetic energy of expansion (\cf\
\citet{PintoE00b} for a discussion and further references). If a
significant luminosity is to be developed, energy must be injected at
later times when the column depth of the ejecta has declined
significantly and radiation can escape.

It is by now widely recognized that this later deposition of energy
results from the radioactive decay of \nifsx.  This isotope is a
by-product of burning to nuclear statistical equilibrium (NSE), the
process which is responsible for liberating much of the energy which
disrupts the star. While $\gamma$-rays from \nifsx decay have yet to
be observed from SNe~Ia, the evidence from the temporal behavior of
the light curve \citep{ColgateMcKee69,ClayColFish69} and from optical
and infrared spectra \citep{Axelrod80,WeaverAW80,KuchnerKPL94} is very
strong. It is further strengthened by the success of the radioactive
decay model in explaining the light curve of SN~1987A
\citep{Xu89,PintoW88a,PintoW88b,PintoWE88}, from which $\gamma$-rays
{\it were} directly observed (\citet{GehrelsLM88} and references
therein). Just as the X- and $\gamma$-ray emission from SN1987A
yielded new and important information about the dynamics of its
explosion, so too observations of SNe~Ia at these energies hold the
promise of significant advances.

Despite 30 years of observation and theoretical study, progress
towards understanding the detailed set of events and the physics
leading to SNe~Ia explosions remains elusive. In particular, the last
decade has seen intense observational efforts in the optical and
near~IR and matching theoretical activity. Though we have learned a
great deal about SNe~Ia, it must still be said that incorporating
current data into a tight theoretical picture remains an enterprise
fraught with difficulty. SNe~Ia explosion models are frustrated by
uncertainties about thermonuclear flame physics and progenitor
evolution, while the interpretation of optical data is hindered by
uncertainties in the non-LTE atomic physics of nearly-neutral
silicon-, and especially, iron-group ions.

If there exists a ``standard model'' for SNe Ia, the current definition
would be a C-O WD in a close binary system which is pushed by accretion
very near the Chandrasekhar mass limit, 1.38 \Msun. As the central density
of the dwarf rises above $10^9$~\gcc, carbon ignites near the center,
leading to an outward-propagating thermonuclear flame. This flame is
Rayleigh-Taylor unstable and very quickly becomes turbulent, leading to a
marked acceleration in its progress, as necessary to achieve an energetic
explosion (\cf\ \citet{WoosleyW86}). This turbulence also makes direct
numerical simulation intractable (\citet{NiemeyerW97,KhokhlovOW97}, and
references therein); current models are thus hampered by a lack of
predictive power.

A possible alternative to the standard model, or perhaps an {\it
addition} to it, is a C-O WD in the mass range $0.6\msun - 0.9 \msun$,
bound to a helium main sequence companion burning helium at it's
center. Accretion in such systems was studied by \citet{LimongiT91}
who found that for accretion rates near $\sim3\times 10^{-8}\
\msunperyear$, of order 0.2~\Msun of helium could be accreted until
this layer detonated near its base.  As others have noted, this is
very near the accretion rate estimated by \citet{IbenT91} who found
this rate, driven by gravitational wave radiation, to be insensitive
to the mass ratio of the two components, thereby alleviating the need
to fine-tune progenitor properties to result in such an explosion.

The possibility that this ignition mechanism might lead to a Type~Ia
supernova was first suggested by \citet{Livne90}, and studied in 2-D
by \citet{LivneG91}.  In one dimensional models, detonation in the
helium layer produces an inward moving, focused compression wave which
drives the central density above $10^8$~\gcc and temperature above
$10^9$~K, causing central carbon ignition under explosive
conditions. The subsequent evolution and nucleosynthesis was studied
by \citet{WoosleyW94} in 1-D, and by \citet{LivneA95} in 2-D. While
the progress of the explosion is rather different in 2-D, both
calculations produce very similar results which may indicate the
robustness of this mechanism. The models possess a number of
attractive properties, such as production of $0.1-0.9$~\Msun of
\nifsx, with the remainder of the original CO WD going to silicon
group isotopes. As Woosley \& Weaver have pointed out, such sub-\Mch\
models may be the production sites for ${}^{44}$Ca (produced as
${}^{44}$Ti), and ${}^{48}$Ti (made as ${}^{48}$Cr), neither of which
are accounted for by either Type~II supernovae or \Mch SNe~Ia
\citep{TimmesWW95}. Even more attractive is the fact that, since
burning takes place at lower densities than in \Mch stars, these
models do not suffer from the problem of excess electron capture and
the resultant over-production of rare neutron-rich species, such as
${}^{54}$Fe, ${}^{58}$Fe, ${}^{54}$Cr, and ${}^{58}$Ni which plague
\Mch models (though accretion at extremely high rates accompanied by
strong winds may be able to cure this, \cf \citet{Brachwitzetal00}).
The amount of \nifsx produced increases monotonically with the mass of
the WD, implying greater maximum brightness for more massive
stars. The increased mass may also mean a longer diffusion time, and
the correlation of both these properties is in the right direction to
explain the maximum brightness-decline rate relationship described by
\citet{Phillips93}. There is some hint of this in the bolometric light
curves computed by Woosley \& Weaver and by Livne \& Arnett, but a
more accurate comparison with observations requires multigroup
transport with velocity broadened line opacities as in
\citet{PintoE00b}.

It is not known what fraction, if any, of observed SNe~Ia are due to
sub-\Mch explosions. \citet{IbenT91} estimated the rate for He
symbiotic systems to explode at the rate of 1 per century in the
galaxy, which compares well with the SNe~Ia rate determined by
\citet{VandenberghT91}, \citet{Cappelaroetal97}, and
\citet{HamuyP99}. In principle, it ought to be possible to distinguish
\Mch and sub-\Mch explosion models from their predicted optical
light curve and spectral evolution. Years of tuning \Mch models to
achieve a match between observed and computed spectra have resulted in
a defining list of properties which the successful Ia explosion must
possess.  ``Successful'' models include the \citet{NomotoTY84} Model
W7 \citep{Harkness89} and the \citet{WoosleyW91} Model DD4
\citep{Kirshneretal93,Eastman95,Pinto95,PintoE00b}.  Sub-\Mch models
have, on the other hand, remained largely unexplored, though some
sub-\Mch models share many of the same desirable properties as their
more massive cousins. The spectra of current sub-\Mch models reported
in the literature are too blue \citep{NugentBBFH97,Hoflichetal96}
resulting in part from the presence of iron and radioactivity in the
outer layers. They also have not produced, to date, Ca at sufficiently
high velocities to match a typical SNe~Ia.  We feel these results are
far from conclusive, however.  Models for sub-\Mch explosions have not
yet undergone the same degree of tuning as their more massive cousins
to bring them into better agreement with observation.  Current
spectrum and light curve calculations are rendered uncertain by their
lack of time dependence, especially in the high velocity, low density
surface layers.

Since the sub-\Mch models have less mass (\eg~0.8 versus 1.38 \Msun),
they might be expected to reach maximum light in less time than a
\Mch explosion. Estimates by \citet{ContardoL98} and
\citet{Riessetal99b} give rise times for nearby SNe~Ia in the range 19
to 23 days in B.  It would be nice if radiation transport simulations
of the light curve evolution were accurate enough to conclusively rule
out sub-\Mch models (or \Mch models) by comparison with
observations. Unfortunately, the effective opacities, and even much of
the basic physics, used in most calculations performed to date remain
highly uncertain.  To be confident in estimates of the rise time, one
requires knowledge of the UV line opacity of low-ionization nickel and
cobalt -- opacities accurate to a factor much smaller than the mass
difference between \Mch and sub-\Mch SNe models.  Errors in the
opacity, which comes principally from velocity-broadened UV lines of
nickel, cobalt, and iron, translate directly into errors in the rise
time and peak brightness \citep{PintoE00a,PintoE00b}.  Most SNe~Ia
light curve calculations (\eg~\cite{HoflichMK93}; \cite{PintoE00b})
have been based solely on lines from the Kurucz list \citep{Kurucz91},
which likely underestimates the nickel and cobalt opacities
significantly. We have compared Rosseland mean opacity values computed
for a pure cobalt composition at $\rho=10^{-12}$~\gcc and $T=20,000$~K
using the Kurucz list and transition data from the OPAL opacity code
\citep{IglesiasRW90,IglesiasRW92} and find that the OPAL opacity value
is approximately 5 times larger.  Thus, in the \Mch calculations by
\citet{Hoflichetal96}, the model with the slowest rise time still
reaches bolometric maximum light in only 15 days.  In a light curve
calculation by \citet{PintoE00b} of a similar model, the predicted
bolometric maximum light was at 15 days, even though the $B$ and $V$
light curves did not peak until 20 days. However, the available
evidence is that the time of bolometric maximum corresponds to the
time of $B$ maximum.  Such discrepancies are enough to raise questions
about the accuracy of current calculations and to doubt claims that
sub-\Mch models are, at present, quite ruled out on theoretical
grounds.

In this paper we present a simple test of the sub-\Mch model which is
quite insensitive to most details of either the modeling or the exact
nature of the supernov\ae\ themselves. It is based upon the fact that
the surface helium detonation in most sub-\Mch models produces
significant yields of \nifsx at high velocities which are absent in
the \Mch models.  High energy photons produced by the decay of \nifsx
can therefore escape largely unimpeded from these surface layers,
while photons released by decay in the radioactive core are strongly
attenuated. A lack of observed high-energy emission at early times
from SNe~Ia would thus argue strongly, and probably fatally, against
the sub-\Mch model. On the other hand, detection of early emission at
high energies would argue strongly in {\sl favor} of these models, as
it is difficult to produce significant quantities of radioactivity in
the surface layers of \Mch explosions. (While hydrodynamic mixing in
pure C-O models might conceivably lead to significant \nifsx at high
velocities as well, the compositional stratification deduced from
early-time spectra would be destroyed by such a process.)  As we shall
show, the decay of surface \nifsx in sub-\Mch explosions produces
considerable Fe, Co and Ni K$\alpha$ emission between 6 and 8 keV, at
flux levels great enough to be detected from extragalactic supernovae,
possibly by current and upcoming missions, and almost certainly by the
proposed Con-X Observatory. They also emit \nifsx decay lines which,
due to the 6.1 day half-life of \nifsx, peak earlier and at higher
luminosities than in \Mch models.

Most aspects of $\gamma$-ray transport in SNe~Ia (and supernovae in
general) have been thoroughly explored and reported on elsewhere.  The
$\gamma$-ray and hard (Compton scattering) X-ray continuum evolution
of typical \Mch explosion model was discussed by \citet{GehrelsLM87},
and especially by \citet{BurrowsT90} and \citet{BurrowsSR91}.
\citet{ClaytonT91} investigated $\gamma$-ray transport in \Mch SNe Ia
models, specifically W7, and discussed the importance of
bremsstrahlung emission in forming the keV X-ray
continuum. \citet{HoflichWK98} presented results of Monte Carlo
$\gamma$-ray transport calculations for both \Mch and sub-\Mch
explosion models, and pointed out differences in \nifsx $\gamma$-ray
line light curve evolution (see section~3).  However, the present work
is the first to describe the X-ray properties of sub-\Mch explosion
models, and to propose observations which clearly discriminate between
explosion models.  In addition, there are significant differences
between our $\gamma$-ray results for sub-\Mch models and those
obtained by \citet{HoflichWK98} which we shall describe below.

The remainder of this paper is organized as follows: in section 2 we
briefly describe the methods used for our calculations. In section 3
we summarize properties of the models we have investigated -- more
detailed descriptions can be found in the original references -- and
present our results, describing and comparing the X-ray and
$\gamma$-ray spectral evolution of \Mch and sub-\Mch models. In
section~4 we discuss prospects for positively detecting the unique
properties of a sub-\Mch event with current and future X- and
$\gamma$-ray missions. In section~5 we describe the evolution of
$\gamma$-ray line profiles and what we might learn with sufficient
sensitivity.  Section~6 gives a short summary of the results.

\section{Computational Methods}

The transport of nuclear decay $\gamma$-rays was computed using an
updated version of the Monte Carlo (MC) $\gamma$-ray transport code
FASTGAM \citep{PintoW87}. FASTGAM includes pair production and
photoelectric opacities and incorporates Compton scattering in the
limit of zero electron temperature. It follows photons emitted by
nuclear decay, pair annihilation, and fluorescence following K-, L-,
and M-shell vacancies (induced both by photoionization and electron
capture). These photons are followed until they escape from the
supernova or are destroyed by absorption.  FASTGAM computes the energy
deposition which results from these processes, the primary electron
energy spectrum produced by Compton recoil, and the emergent $\gamma$-
ray spectrum.

Because the supernova ejecta are ionized, these primary electrons, with
kinetic energies ranging up to $\sim1$~Mev, lose energy primarily by
exciting collective plasma oscillations. Other loss mechanisms include
atomic ionization and excitation, and, as first noted by
\cite{ClaytonT91}, they also copiously produce bremsstrahlung X-rays.
Under certain conditions which we discuss below, these X-rays dominate
the continuum emission from $\sim1$ to $\sim50$~keV. 

Because the transport equation is linear in the emissivity, the bremsstrahlung
spectrum can be calculated separately from the MC code and then
added linearly to the primary $\gamma$-ray spectrum calculated by FASTGAM.  
Therefore, following the MC calculation, the bremsstrahlung contribution 
was computed
deterministically using a 1-D, spherical, multi-frequency comoving frame
transport code \citep{EastmanP93}. The opacity is dominated by K- and
L-shell photoionization, followed by electron scattering.  The spectrum of
primary Compton electrons, $S(E)$ (e$^-$~s$^{-1}$~erg$^{-1}$~atom$^{-1}$)
resulting from the MC calculation is used to compute the bremsstrahlung
emissivity and solve for the emergent X-ray continuum flux.

The bremsstrahlung emissivity was computed using the {\it continuous slowing
down approximation}, given by
\begin{equation}
\eta_\nu = {h\nu\over 4\pi}\int_{h\nu}^\infty dE S(E) \sum_Z N_Z
\int_{h\nu}^E 
\frac{ \partial\sigma_Z(E^\prime)/\partial\nu }{L(E^\prime)}
\,dE^\prime\ \ \ \ 
[{\rm ergs\ cm^{-3}\ s^{-1}\ Hz^{-1}\ ster^{-1}}]
\label{brememis}
\end{equation}
where $L(E)$ is the {\sl loss function}: the e-folding path length for the
electron's energy (\cf\ \citet{Axelrod80}). The partial bremsstrahlung cross
section, $\partial\sigma_Z(E^\prime)/\partial\nu$, for an electron of
energy $E$ to produce a photon of frequency $\nu$ while interacting with a
nucleus of charge $Z$, was taken from the numerical calculations of
\citet{KisselMP81}.  The output of the deterministic transport calculation
is the monochromatic photon energy density as a function of depth through
the gas and the emergent X-ray flux as measured in the observer frame.

The X-ray continuum from bremsstrahlung emission is capable of
producing additional K-shell vacancies. For instance, photons with
energy $> 7.117$~keV are capable of ionizing iron. These additional
vacancies contribute to the K$\alpha$ production rate, and were not
included in the MC calculation. A further source of vacancies comes
from direct impact ionization of the K-shell by non-thermal electrons,
however \citet{TheBC94} have shown that the collisional ionization
contribution is negligible at early times.  The photoionization
contribution to the K-shell vacancy rate was included as follows: the
monochromatic photon energy density obtained from the deterministic
transport solution was used to compute the additional K-shell
photoionization rate. These were combined with fluorescence yields
\citep{KaastraM93} for K$\alpha$ and K$\beta$ emission following a
K-shell vacancy to obtain the line emissivities. These were then used
to perform another deterministic transport calculation (the opacity is
the same as for the bremsstrahlung calculation), giving the emergent
flux, in the observer frame, of K-shell lines produced from
photoionization by bremsstrahlung X-rays; it is at most a few percent
contribution. 

To summarize, the composite spectrum is computed in three steps; 1)
the transport of decay $\gamma$-rays is calculated using a Monte-Carlo
code which gives the emergent spectrum of unscattered and
down-scattered $\gamma$-rays, the K-shell line flux, and the spectrum
of primary, Compton scattered electrons. 2) We use the {\it continuous
slowing down} approximation to compute the bremsstrahlung emissivity,
which is plugged into a deterministic transport code to obtain the
monochromatic X-ray photon energy density and the emergent
bremsstrahlung x-ray spectrum. 3) Additional K-shell line production
produced when bremsstrahlung X-rays photoionize $1s$ electrons, is
accounted for in a final step. All that is needed for this is the
photonionization rate, which is computed using the monochromatic X-ray
photon energy density obtained in step 2. The emergent spectra
obtained from these three steps are added together to produce the
final, composite spectrum.

The inner-shell photoionization rate due to X-rays is small compared
with the rate from primary $\gamma$-rays. The loss function is
determined primarily by the ionization state of the gas, and this is
determined in turn by the balance of valance-shell photoionization by
UV photons and radiative recombination. Thus, a fully self-consistent
calculation would require a complete solution of the radiation
transport and statistical equilibrium at all energies from the
$\gamma$-ray to the infrared.  For a gas with free (thermal) electron
density $n_e \gtrsim n_{ion}$, however, the loss function is
insensitive to ionization, a condition which the supernova ejecta
satisfy throughout their evolution. Our three-step procedure, MC
$\gamma$-ray transport, bremsstrahlung from primary and secondary
electrons, and additional K-shell lines from X-ray photoionization,
can thus be expected to closely approximate a full, self-consistent
solution. 

\begin{figure}[!t]
\epsscale{0.7}
\plotone{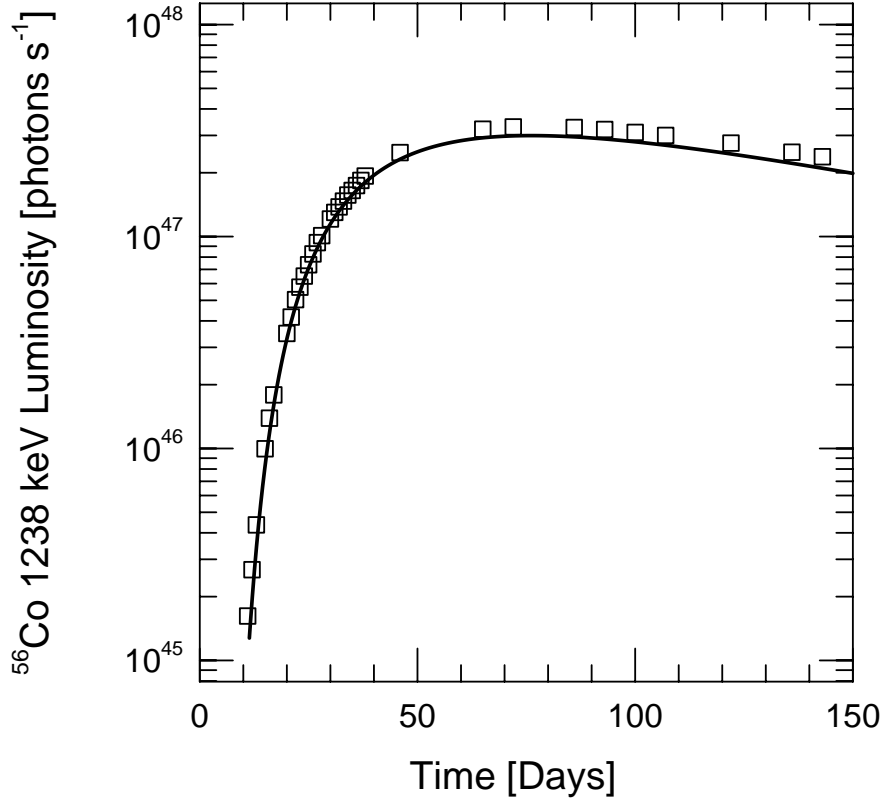}
\caption{Comparison of the \cofsx\ 1238 keV line light curve for Model
W7, as computed with the Monte Carlo code FASTGAM (squares), to that
computed using the deterministic transport method described in section~2.}
\label{mcdetermcomp}
\end{figure}

The emergent spectrum computed by FASTGAM consists primarily of
velocity broadened $\gamma$-ray decay lines superimposed upon a
smooth, Compton scattering continuum. Sampling the emergent intensity
with enough energy resolution to produce accurate line fluxes requires
a large expenditure of computing time. As the opacity for $\gamma$-ray
lines is absorptive (Compton scattering {\em into} the line profile,
while included in the MC treatment, is negligible), a simpler
procedure is to compute individual line fluxes directly using a
deterministic transport algorithm.  The procedure for this is
straightforward and was described in \citet{PintoW88b} and
\citet{EastmanWWP94} (we note however that that in the present work we
have used $\theta_{sc}=56^\circ$ for the mean scattering angle). This
deterministic approach to computing $\gamma$-ray line transport is
approximate, less accurate than MC for computing energy deposition,
and does not give the Compton continuum; it does do a fine job,
however, at correctly predicting the $\nu$-integrated emergent line
fluxes. This is shown in Figure~\ref{mcdetermcomp}, which compares the
\cofsx\ 1238 keV line light curve for Model W7 computed
deterministically and with FASTGAM.

\section{Results}

The models investigated in this paper consist of two \Mch mass models and
three sub-\Mch models. The properties of these models are summarized in Table~1.

\begin{table}
\begin{tabular}{rcccccccc}
\multicolumn{9}{c}{\bf Table 1: Properties of Explosion Models}	\\
\hline
\hline
name	& $M_{tot}$	& $M_{CO}$	& $M_{He}$ & $E_{51}$ 
& $M({}^{56}{\rm Ni})$	& $M({}^{56}{\rm Ni})$ (surface)
& $M({\rm Si\ group})$	& ref\\
\hline
Model 2 & 0.90	& 0.70	& 0.20	& 0.90	& 0.43	& 0.09	& 0.29	& a \\
M1	& 0.70	& 0.55	& 0.15	& 0.69	& 0.14	& 0.03	& 0.26	& b \\
M8	& 1.10	& 0.90	& 0.20	& 0.25	& 0.71	& 0.17	& 0.25	& b \\
W7	& 1.38	& 1.38	& 0	& 1.2	& 0.63	& 0.0	& 0.29	& c	\\
DD4	& 1.39	& 1.39	& 0	& 1.2	& 0.63	& 0.0	& 0.49	& d	\\

\hline
\hline

\end{tabular}  
\caption{Properties of explosions models considered in this paper. For
a more detailed description, refer to the original references: a)
Woosley \& Weaver 1994; b) Livne \& Arnett 1995; Nomoto, Thielemann \&
Yokoi 1984; d) Woosley \& Weaver 1991.}
\label{modtab}
\end{table}

The \mch models include Model W7 by \citet{NomotoTY84} and Model DD4 by
\citet{WoosleyW91}.  The X-ray and $\gamma$-ray spectral evolution of Model
W7 was studied both by \citet{BurrowsT90} and by \citet{ClaytonT91}.  Model
DD4 is similar in many respects to Model W7, in terms of mass, energy and
total mass of \nifsx\ produced, although in DD4 the flame speed, which in
both models was an adjusted parameter, was allowed to exceed the sound
speed and become a detonation, whereas in Model W7 the flame remained
subsonic. The result is that in W7 only 0.29~\Msun\ of ``silicon group''
nuclei (by which we mean ${}^{24}$Mg through ${}^{40}$Ca) was produced,
whereas in DD4 there is 0.49~\Msun. Despite this, both models have been
shown to give reasonable agreement to spectral observations of maximum
light supernovae \citep{Harkness89,Kirshneretal93}. It is not clear whether
this indicates a lack a sensitivity of such calculations to distinguish
large abundance differences, or a large inherent variation in the objects
themselves.

The sub-\Mch models investigated include Model 2 by \citet{WoosleyW94},
and Models M1 and M8 from \citet{LivneA95}.  Whereas Model 2 was a 1-D
calculation, Models M1 and M8 were 2-D, and have been mapped into a 1-D,
angle averaged structure (courtesy of Professor Arnett).  In terms of the
initial C-O white dwarf mass, the total mass, the amount of accreted
helium, the total amount of \nifsx produced and the final explosion energy,
these three models span a large range of properties.

\begin{figure}[tp]
\epsscale{1.0}
\plotone{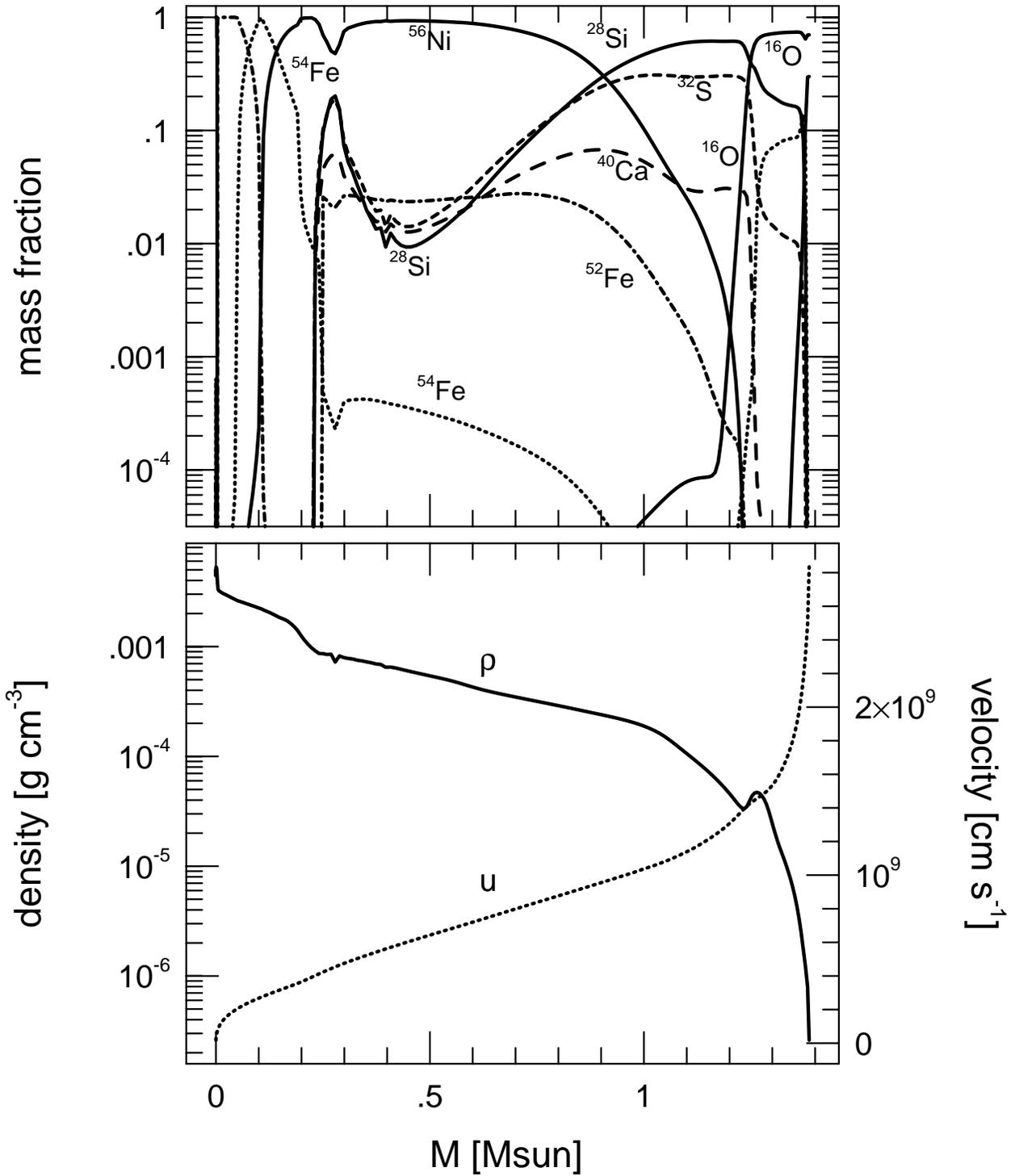}
\caption{Composition, density and velocity structure of the \Mch
delayed detonation Model DD4 (Woosley \& Weaver 1991),
at 1000~seconds after explosion.}
\label{edefmod}
\end{figure}

\begin{figure}[tp]
\epsscale{1.0}
\plotone{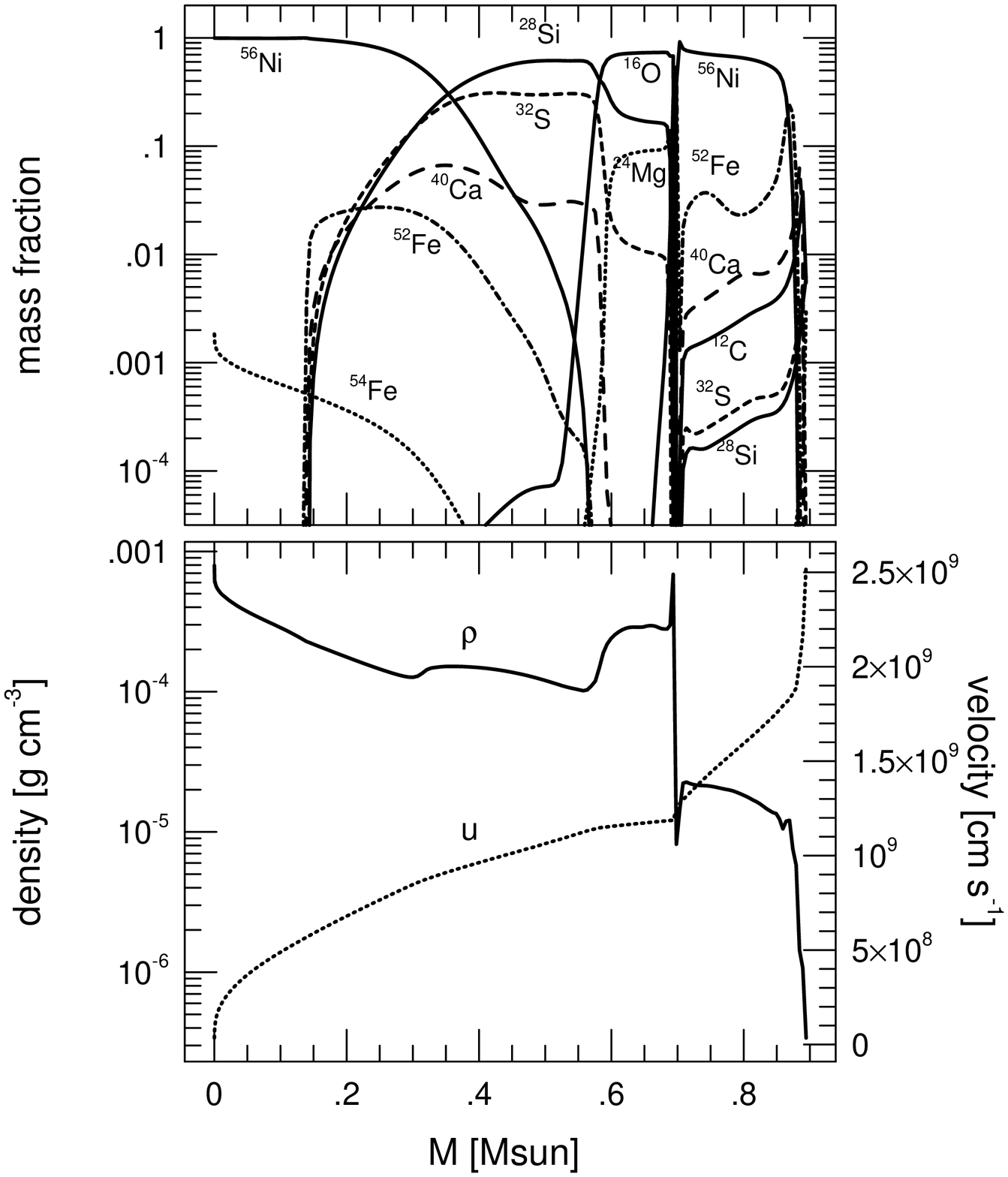}
\caption{Composition, density and velocity structure of the sub-\Mch,
edge-lit helium detonation Model 2 (Woosley \& Weaver 1994), at
1000~seconds after explosion.}
\label{hedtb11mod}
\end{figure}

Figure~\ref{edefmod} shows the composition, density and velocity structure
of Model DD4, which may be compared to that of Model 2, shown in
Figure~\ref{hedtb11mod}. The biggest difference between the two classes of
models, for present purposes, is the large mass fraction of \nifsx\ on the
surface of the sub-\Mch models.

\begin{figure}[tp]
\epsscale{1.0}
\plotone{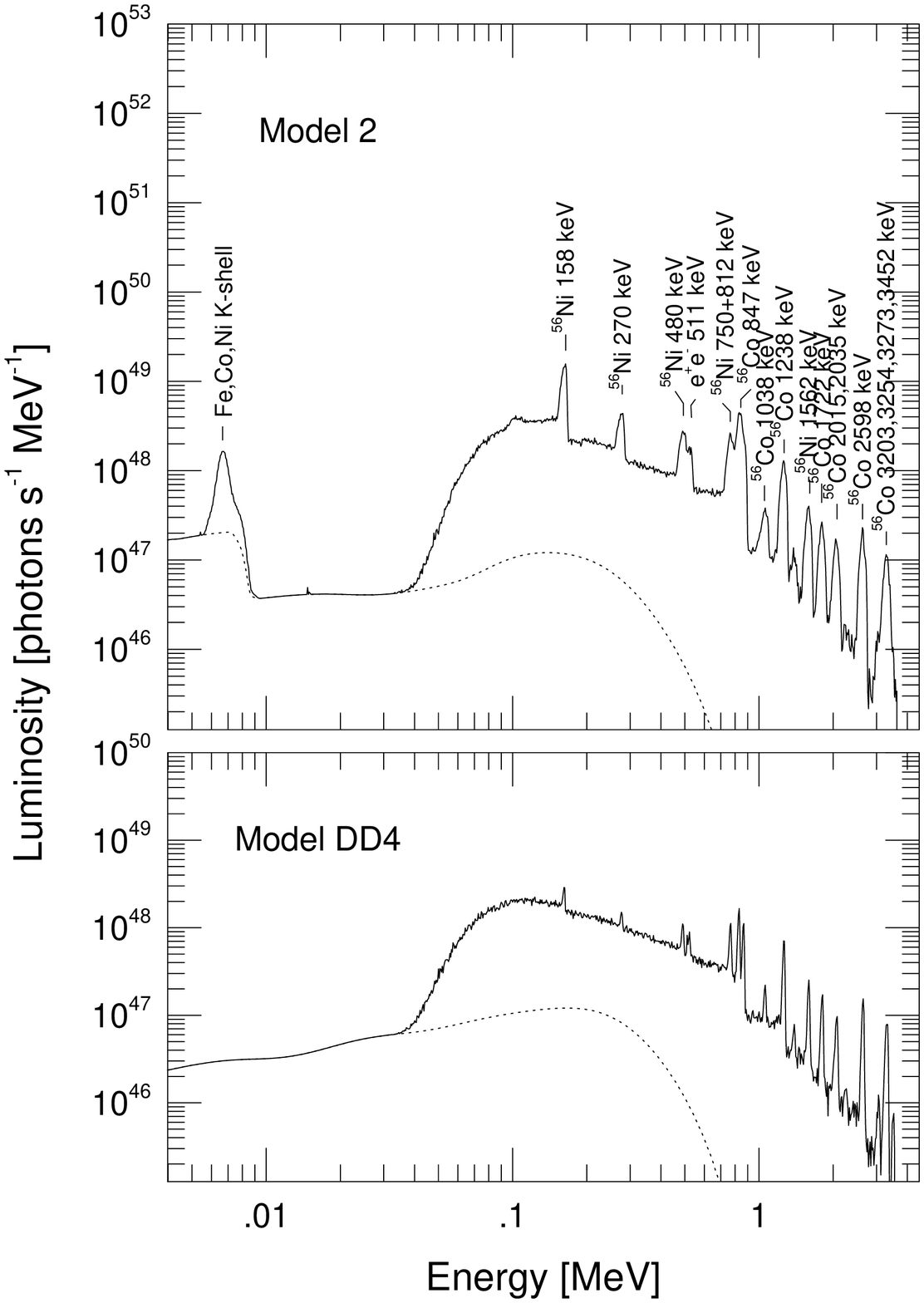}
\caption{Comparison of the 20 day spectra of Model 2 and DD4. The
dotted line shows the contribution to the spectrum from bremsstrahlung
emission.}
\label{chsubchcomp}
\end{figure}

Figure~\ref{chsubchcomp} compares the 20 day spectra of Models 2 and DD4.
One consequence of the surface \nifsx in the sub-\Mch models is the strong
\ka line at 7~keV. This feature is completely absent from \Mch models.  In
Model~2 (and the other sub-\Mch models), K-shell line emission is
superimposed upon a bright bremsstrahlung continuum, dominating the
spectrum near 7~keV. At 20 days this feature is a combination of
contributions from Ni (10~percent) Co (73~percent) and Fe (16~percent),
reflecting the relative abundances of these ions at the time.  The energies
and fluorescence yields for the 12 contributing lines, taken from
\citet{KaastraM93}, are summarized in Table~2. For all three elements,
roughly 30~percent of the emission is due to the K$\alpha_2$ transition,
59~percent to K$\alpha_1$, 4~percent to K$\beta_3$ and the remaining
7~percent to K$\beta_1$. The effect of the continuum edges of the Fe, Co
and Ni K-shell photoionization crossections near $E\gtrsim7$~keV is clearly
evident in the Model~2 spectrum. The jump is strong in Model~2 because the
the outer material is rich in Fe peak elements. No such jump is present in
the spectrum of Model~DD4, because the Fe abundance is much lower in the
outer part of the ejecta, dominated as it is by silicon group elements
and unburned carbon and oxygen.

\begin{table}
\begin{tabular}{rcc}
\multicolumn{3}{c}{\bf Table 2: K-shell Lines}	\\
\hline
\hline
transition	& $E$ [keV]	& yield	\\
\hline
Fe K$\alpha_2$	& 6.3915	& 0.1013 \\
Fe K$\alpha_1$	& 6.4047	& 0.2026 \\
Fe K$\beta_3$	& 7.0567	& 0.0127 \\
Fe K$\beta_1$	& 7.0583	& 0.0254 \\
Co K$\alpha_2$	& 6.9151	& 0.1084 \\
Co K$\alpha_1$	& 6.9295	& 0.2168 \\
Co K$\beta_3$	& 7.6472	& 0.0136 \\
Co K$\beta_1$	& 7.6489	& 0.0272 \\
Ni K$\alpha_2$	& 7.4611	& 0.1226 \\
Ni K$\alpha_1$	& 7.4782	& 0.2451 \\
Ni K$\beta_3$	& 8.2623	& 0.0154 \\
Ni K$\beta_1$	& 8.2642	& 0.0309 \\
\hline
\hline
\end{tabular}  
\caption{Fe, Co and Ni K-shell lines produced by sub-\Mch SNe~Ia
(taken from Kaastra \& Mewe 1993).}
\label{kshelllines}
\end{table}

\begin{figure}[!t]
\epsscale{0.7}
\plotone{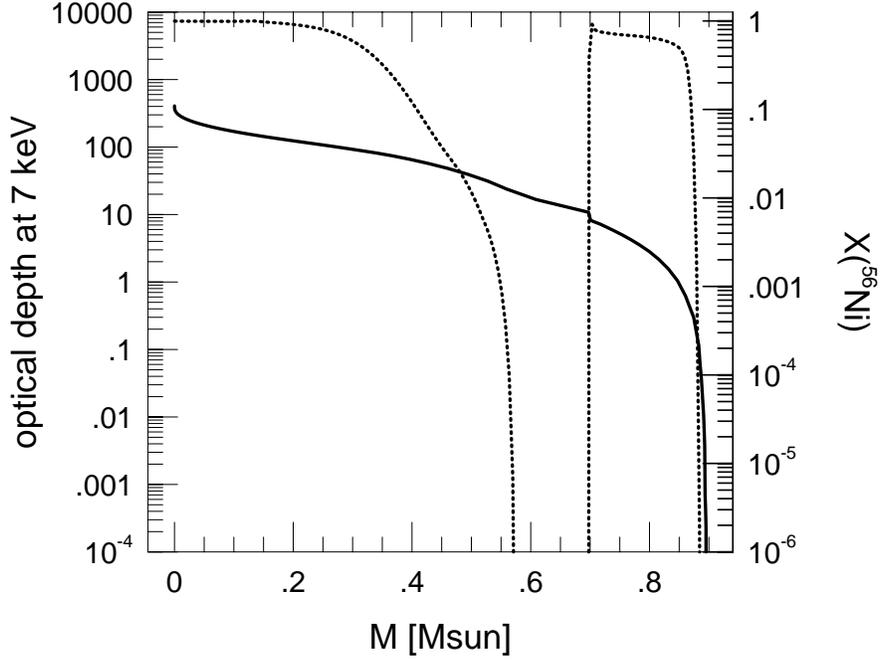}
\caption{This figure shows the optical depth at 7 keV (solid line) 
and \nifsx mass fraction (dotted line), versus mass, in Model~2 at 100
days after explosion. Only the gas situated at optical depth of order
unity contributes to the emergent K-shell line flux.}
\label{tauhedt}
\end{figure}

\begin{figure}[!t]
\plotone{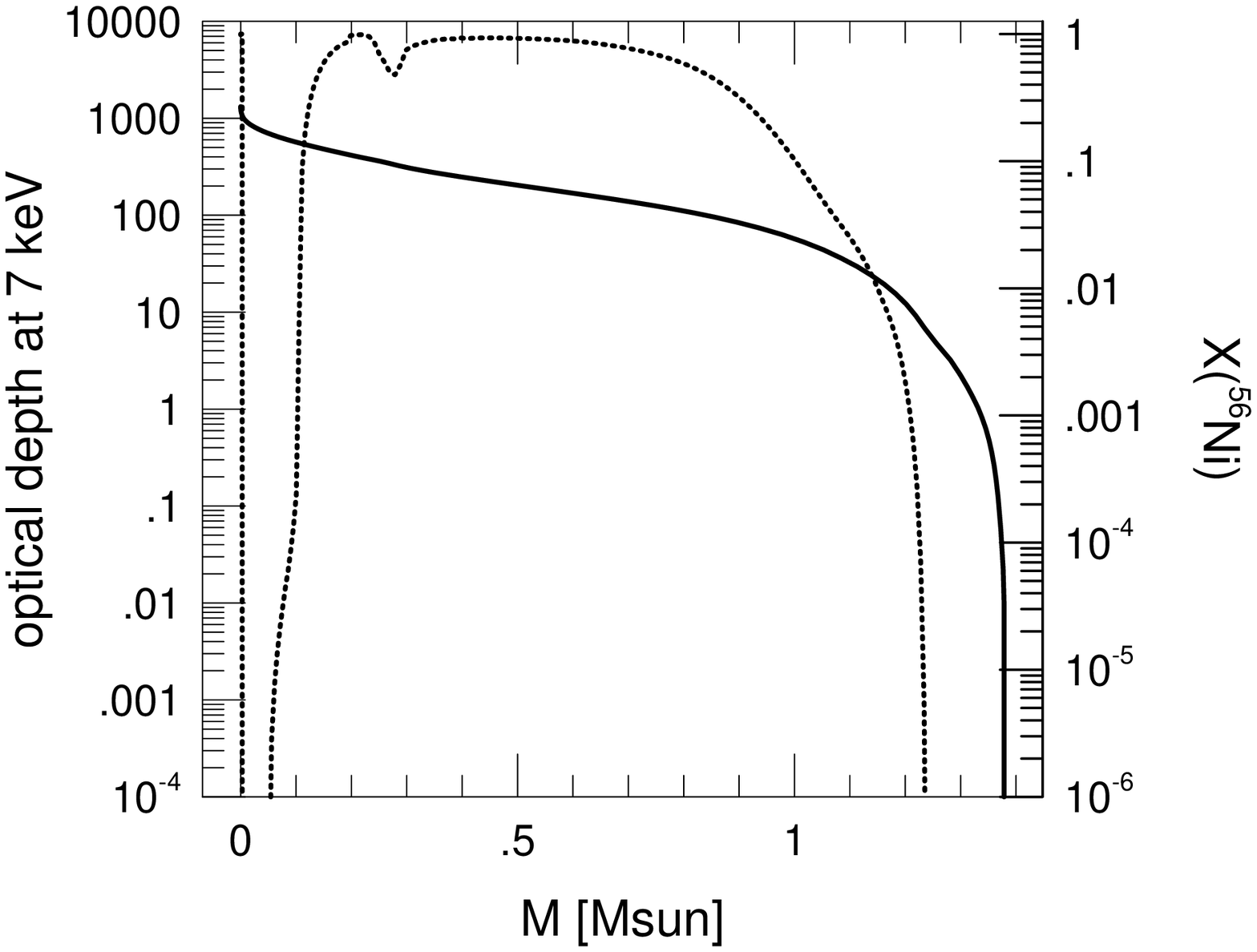}
\caption{Same as Figure \ref{tauhedt}, but for Model DD4. In this case
the \nifsx (\cofsx and \fefsx at 100 days) sits at large optical
depth, and K-shell photons produced in the core are absorbed before
they are able to reach the surface.}
\label{tauedef}
\end{figure}

Although both Models DD4 and W7 have 0.63~\Msun of \nifsx, and
therefore produce a large number of \ka photons, the optical depth at
7~keV remains large enough to absorb the X-ray emission from the
\nifsx core until several hundred days after explosion. This is shown
by Figures~\ref{tauhedt} and \ref{tauedef}, which compare the optical
depth at 7~keV to the \nifsx mass fraction in Models 2 and DD4,
respectively, at 100 days. As the opacity is not affected by
ionization (for low mean ionization), the optical depth at other times
can be obtained by scaling as $\tau\propto t^{-2}$. Most of the
optical depth at 7~keV is due to L-shell photoionization.  Only
emission from material at $\tau \lesssim 1$ is able to escape the
surface. Consequently, the emergent K-shell line emission is due
entirely to the presence of \nifsx on the surface.

\begin{figure}[!t]
\epsscale{1.0}
\plotone{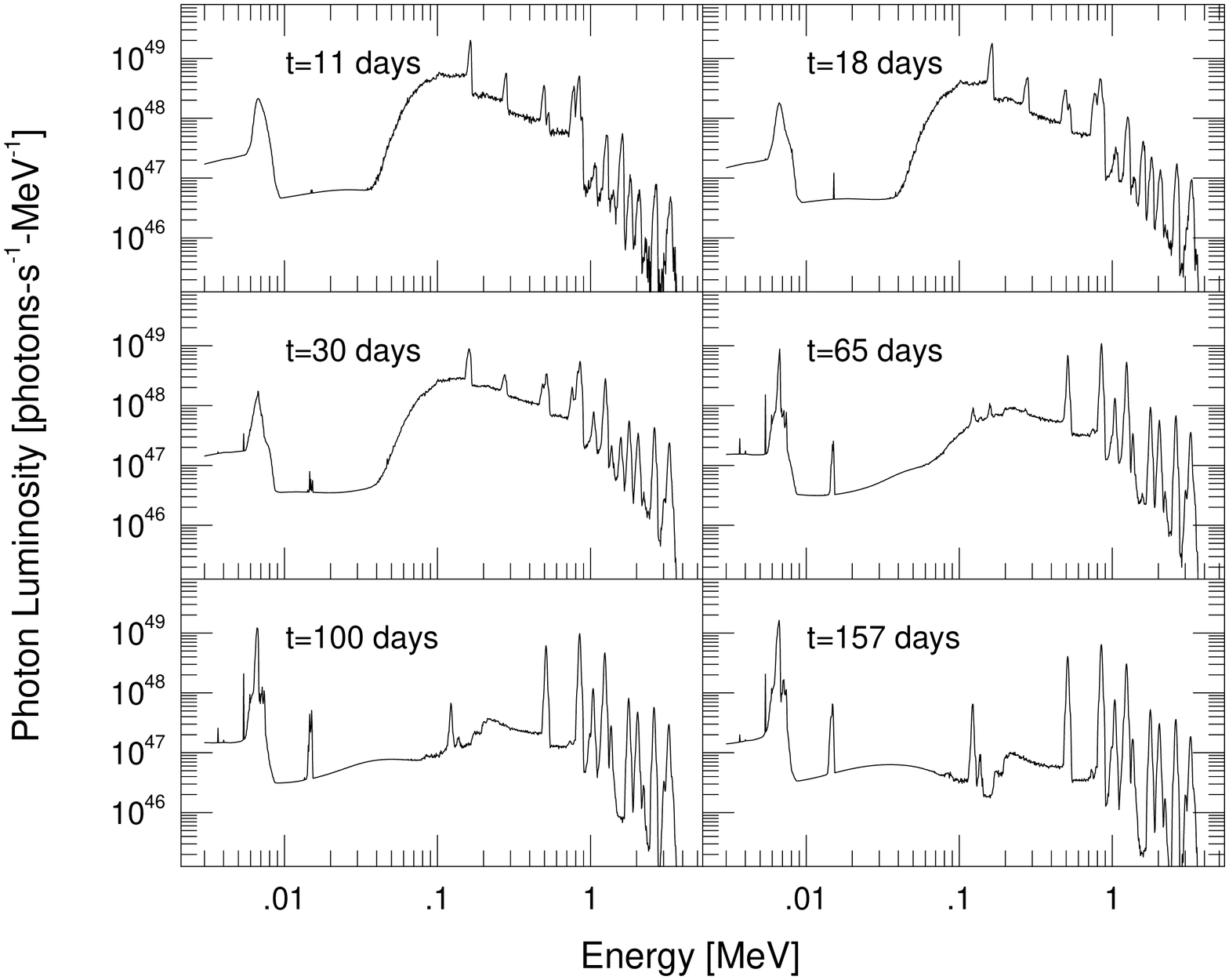}
\caption{Spectral evolution of Model 2.}
\label{hedtevol}
\end{figure}

\begin{figure}[!t]
\epsscale{1.0}
\plotone{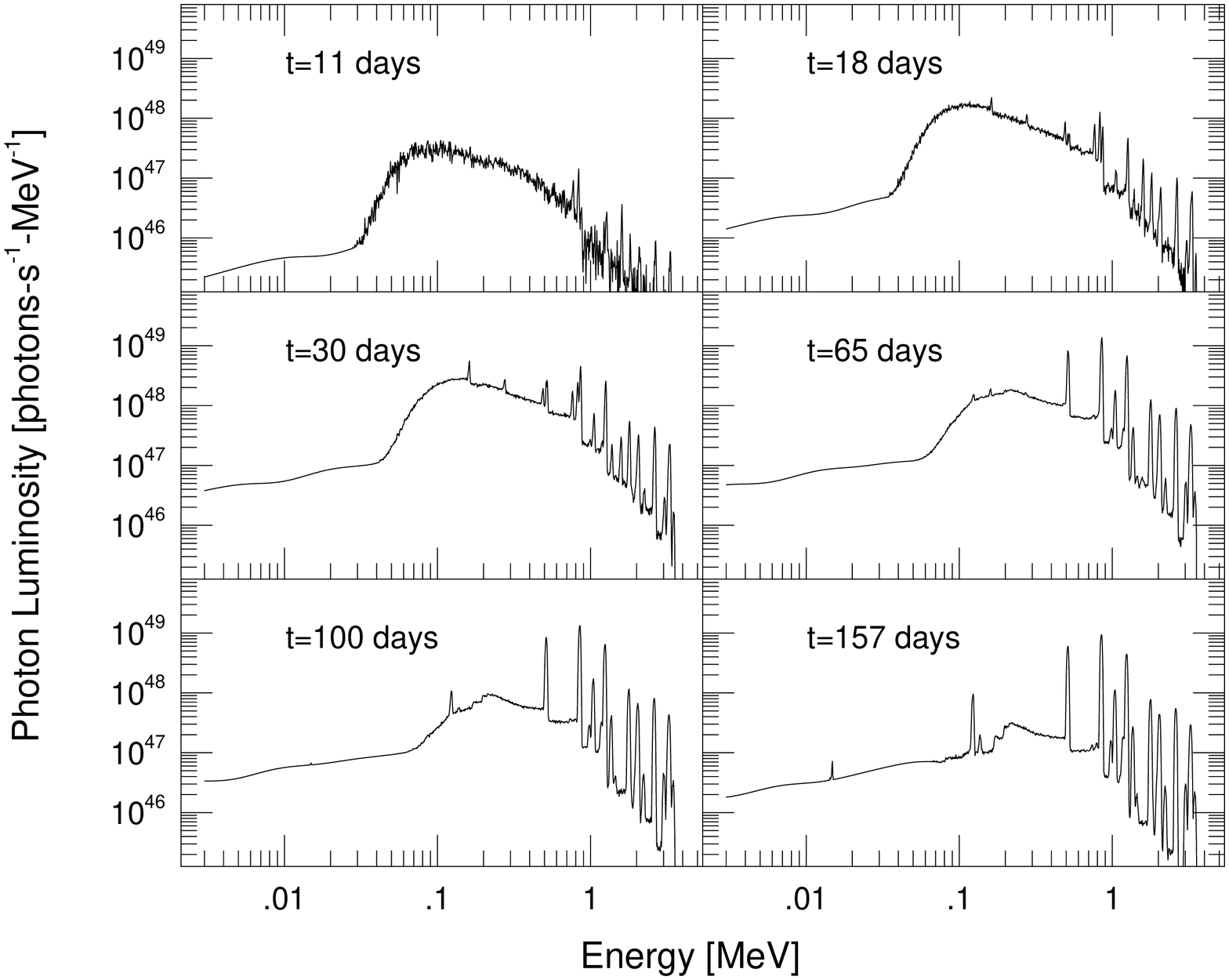}
\caption{Spectral evolution of Model DD4.}
\label{edefevol}
\end{figure}

Spectra from Model~2 at various times during the first 160 days are shown
in Figure~\ref{hedtevol}. These should be compared with the spectra of
Model~DD4 at the same times after explosion from Figure~\ref{edefevol}.  At
times $t\lesssim 100$~days, the $E\gtrsim50$~keV spectrum in both cases is
dominated by narrow (see below) $\gamma$-ray decay lines superimposed on a
Compton scattering continuum produced from down-scattered $\gamma$-rays. As
the column density declines with time, the Compton optical depth decreases
below unity and the Compton continuum eventually disappears. An excellent
discussion of the evolution of the Compton continuum is given by
\citet{Xu89} and \citet{XuRM91}.  The smooth continuum at $E\lesssim50$~keV
is due to bremsstrahlung emission.

In Figure~\ref{hedtevol}, the strong feature at 14.4~keV which is visible
in the 100~day spectra and afterwards is a nuclear decay line of
${}^{57}$Co. The abundance of ${}^{57}$Co was assumed to be given by the
solar ${}^{57}\hbox{Fe}/{}^{56}\hbox{Fe}$ ratio (0.027), times the \nifsx
mass fraction in the model. As with the K-shell lines, this line will only
be visible in a sub-\Mch explosion -- the optical depth to core ${}^{57}$Co
is too great for any appreciable escape to occur.

\begin{figure}[!t]
\epsscale{0.7}
\plotone{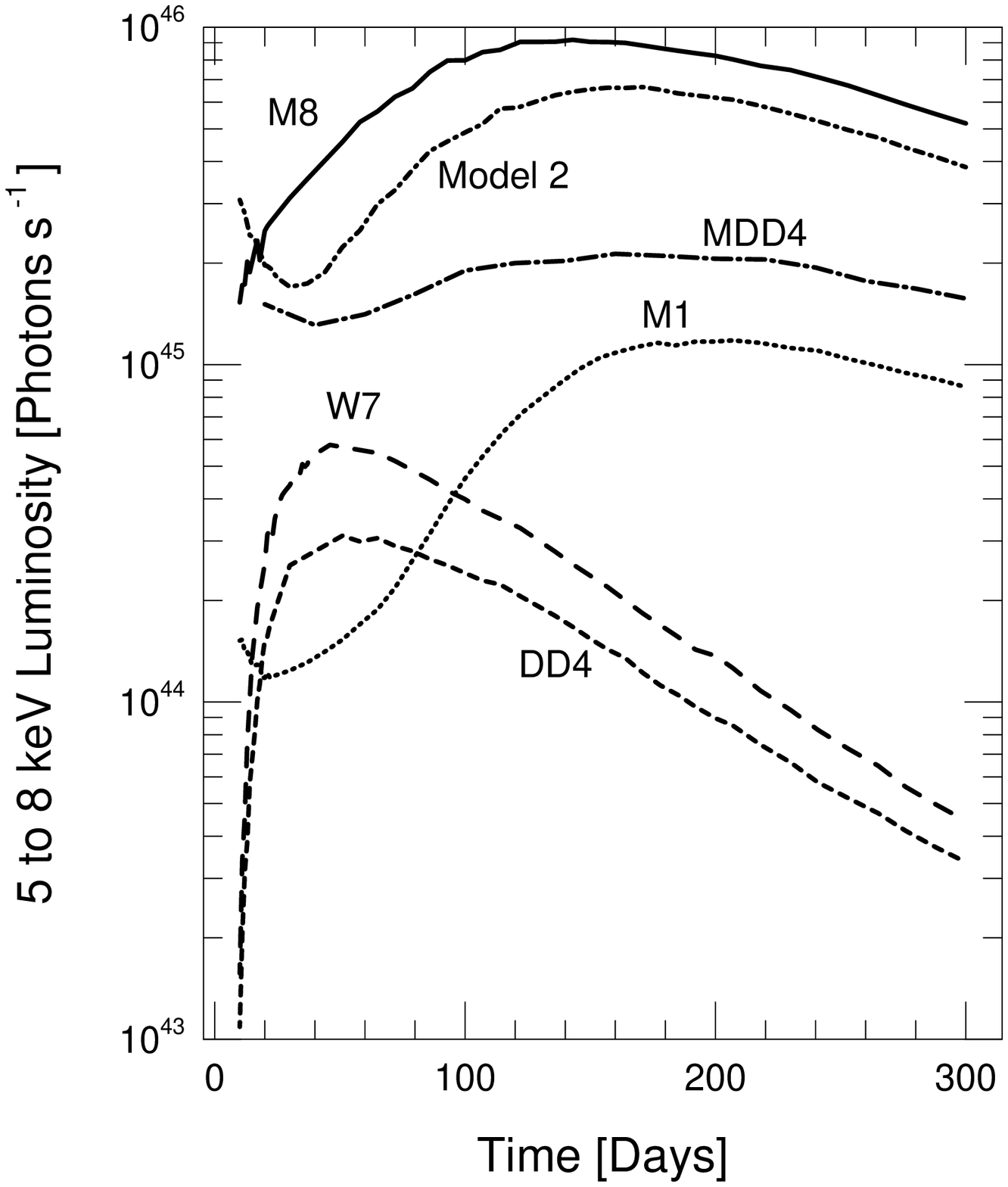}
\caption{Evolution of the integrated K-shell line luminosity (
K$\alpha_2$, K$\alpha_1$, K$\beta_3$, K$\beta_1$).}
\label{kalphaLC}
\end{figure}

The time evolution of the integrated K-shell line luminosity for all models
is shown in Figure~\ref{kalphaLC}. The effect of the surface \nifsx is
dramatic.  Again, because of the large optical depth to core \nifsx, only
\ka emission produced on the surface is able to escape, with the result
that even Model~M1, which has only 22~percent as much \nifsx as Model~W7,
is nearly 100 times brighter in the K$\alpha$ line than W7.

\begin{figure}[!t]
\plotone{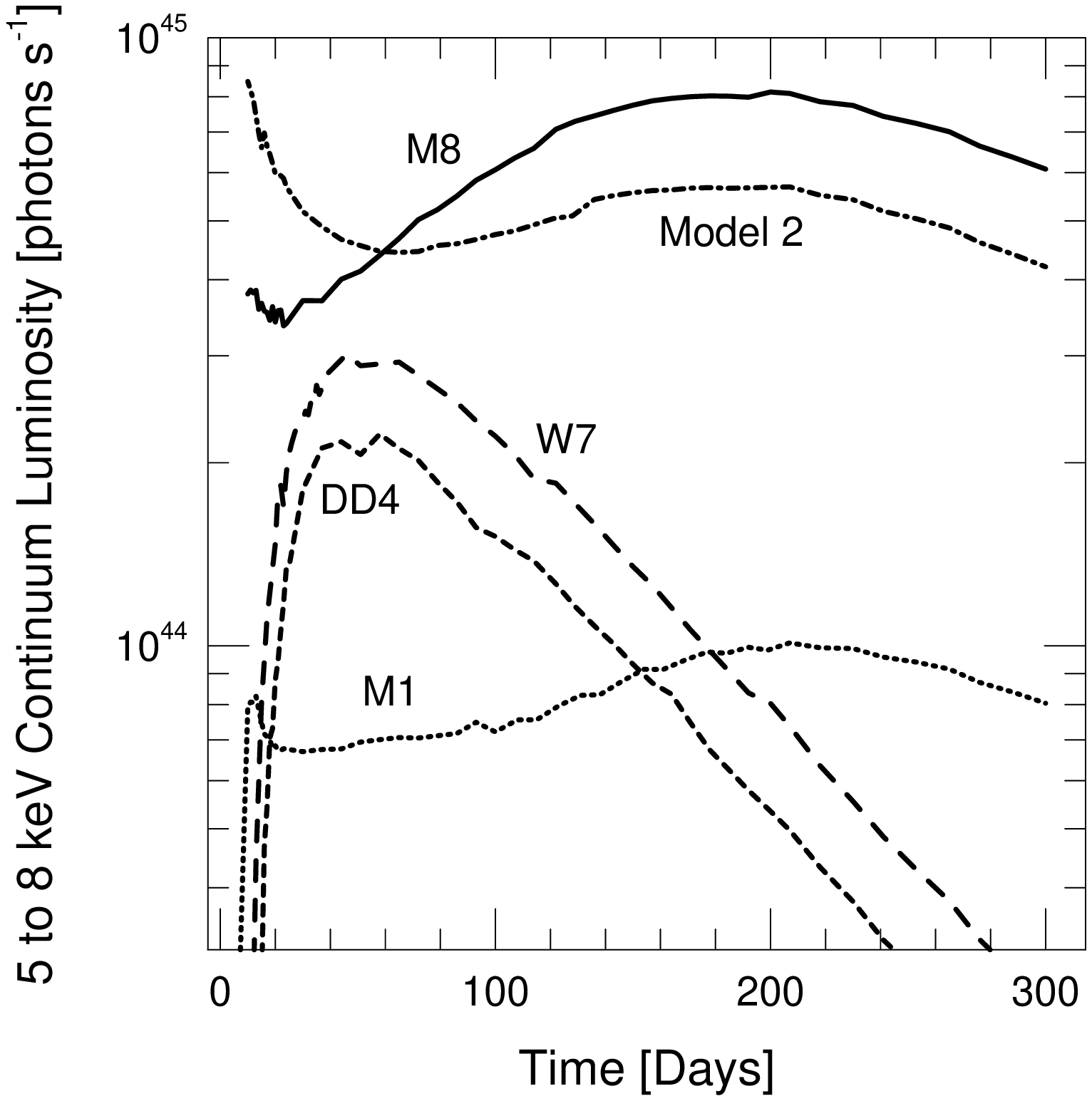}
\caption{Luminosity evolution in the 5 to 8 keV bremsstrahlung
continuum, excluding contributions from K-shell line emission.}
\label{bremLC}
\end{figure}

The bremsstrahlung continuum time evolution behaves somewhat differently
than that of the K-shell lines -- at least in the \Mch
models. Figure~\ref{bremLC} compares the time evolution of the integrated
5~to~8~keV continuum luminosity, excluding contributions from K-shell line
emission, for each of the models.  In the \Mch models, the emergent X-ray
flux must all come from the core. The continuum emission which reaches the
surface is produced at somewhat higher energy, where the K-shell bound-free
opacity is lower. However, the Compton optical depth is not negligible, and
photons are Doppler shifted by the bulk expansion of the gas as they
diffuse out. The amount by which a photon's energy is reduced in diffusing
to the surface can be estimated as
\begin{equation}
{\Delta E\over E}\sim {l\over c t}\sim 
{3 R^2\rho\kappa\over ct}
\end{equation}
where $l$ is the total path length traversed, $R$ is the distance
between point of emission and the surface.
Using $R\approx V  t$ and $\rho\sim 3M/4\pi R^3$, this can be expressed as
\begin{equation}
{\Delta E\over E}\sim
10\left({M\over 1.4 M_\odot}\right)
\left({\kappa\over 0.2}\right)
\left({15\ \hbox{days}\over t}\right)^2
\left({8\times10^8\ \hbox{cm}-\hbox{s}^{-1}\over v}\right)
\end{equation}
It is this Doppler shifting of photons as they diffuse out which
accounts for the fact that there is $E\sim 7$~keV X-ray continuum
emission in \Mch models. The photons are emitted at higher energy,
where the K-shell bound-free optical depth is lower, and therefore
they have an enhanced chance of making it to the surface and
escape. Additionally, since the abundances in the outer layers of the
\Mch models are of lower $Z$, the bound-free optical depth in these
layers is less than in the sub-\Mch case; more X-ray continuum photons
from the core will make it to the surface.

In the sub-\Mch models, the presence of a large iron peak abundance in
the outer layers ensures that the optical depth from K-shell
photoabsorption is very large, preventing core X-rays from emerging at
the surface.  The 5-8~keV X-ray continuum in this case is due only to
surface \nifsx.  For the sub-\Mch models, the controlling factor is
the K-shell bound-free optical depth in the surface layers. Roughly,
\begin{equation}
L_{5-8\ {\rm keV}}\propto
\left[S_{56}(t)\left(1-\exp(-\tau_\gamma)\right)\right]
\exp(-\tau_X),
\label{DepEqn}
\end{equation}
where $S_{56}(t)$ is the $\gamma$-ray emission rate ($[{\rm
ergs\ g^{-1}}]$),
$\tau_\gamma$ is an effective optical depth for trapping
$\gamma$-rays, and $\tau_X$ is the optical depth the 5-8~keV
range. The term in square brackets is a decreasing function of time,
but the $\exp(-\tau_X)$ is an increasing function of time and
dominates the light curve. Thus, for the sub-\Mch models, the flux
increases with time because more of the surface \nifsx is ``exposed''
by the declining absorptive optical depth.  For the \Mch models, we
can write
\begin{equation}
L_{5-8\ {\rm keV}}\propto
\left[
S_{56}(t)\left(1-\exp(-\tau_\gamma)\right)
\right]
\exp(-\tau_{X^\prime}) \int_{\bar E}^\infty \phi(E^\prime)\, dE^\prime
\label{ChXL},
\end{equation}
where $\tau_{X^\prime}$ is the Compton optical depth at X-ray energy,
in the absence of photoabsorption,
$\phi(E)$ is a normalized X-ray emission distribution function, and 
\begin{equation}
{\bar E} \sim E (1 + {3 R^2\rho\kappa\over ct}).
\end{equation}
Equation~\eqn{ChXL} expresses (very approximately) the fact that the
surface flux at energy $E$ results from emission at higher energy in
the interior. At the earliest times, when the Compton optical depth is
high, the Doppler shift incurred by a photon as it random walks its way
out is substantial, and a large fraction of the X-rays are absorbed
before arriving at the surface. As the optical depth drops, there is
less Doppler shift and less absorption, and more of the X-rays can
escape. Once again, the time derivative of the square bracketed term
in Equation~\eqn{ChXL} is negative, while that of the second term is
positive.

It is possible that vigorous hydrodynamic mixing in a \Mch explosion
could bring enough \nifsx up to the surface that the X- and
$\gamma$-ray light curve would more closely resemble that of a
sub-\Mch explosion than a \Mch explosion. As mentioned in the
introduction, we regard such extensive mixing as unlikely because the
compositional stratification deduced from early-time spectra would be
destroyed. However, for the sake of argument, we consider the most
extreme case possible, which is complete homogenization of the
composition of Model DD4. This is shown in Figure~\ref{kalphaLC} as
Model MDD4. Unlike either the unmixed \Mch models or
the sub-\Mch models, the 5~to~8~keV light curve in this case is nearly
flat, and comparable in brightness, at 200 days, to the faintest of
the sub-\Mch models considered, Model~M1. The constancy of the light
curve is evidently due to the deposition term in eqn~\eqn{DepEqn}
decreasing at the same rate that the exponential attenuation term is
increasing.

\begin{figure}[!t]
\plotone{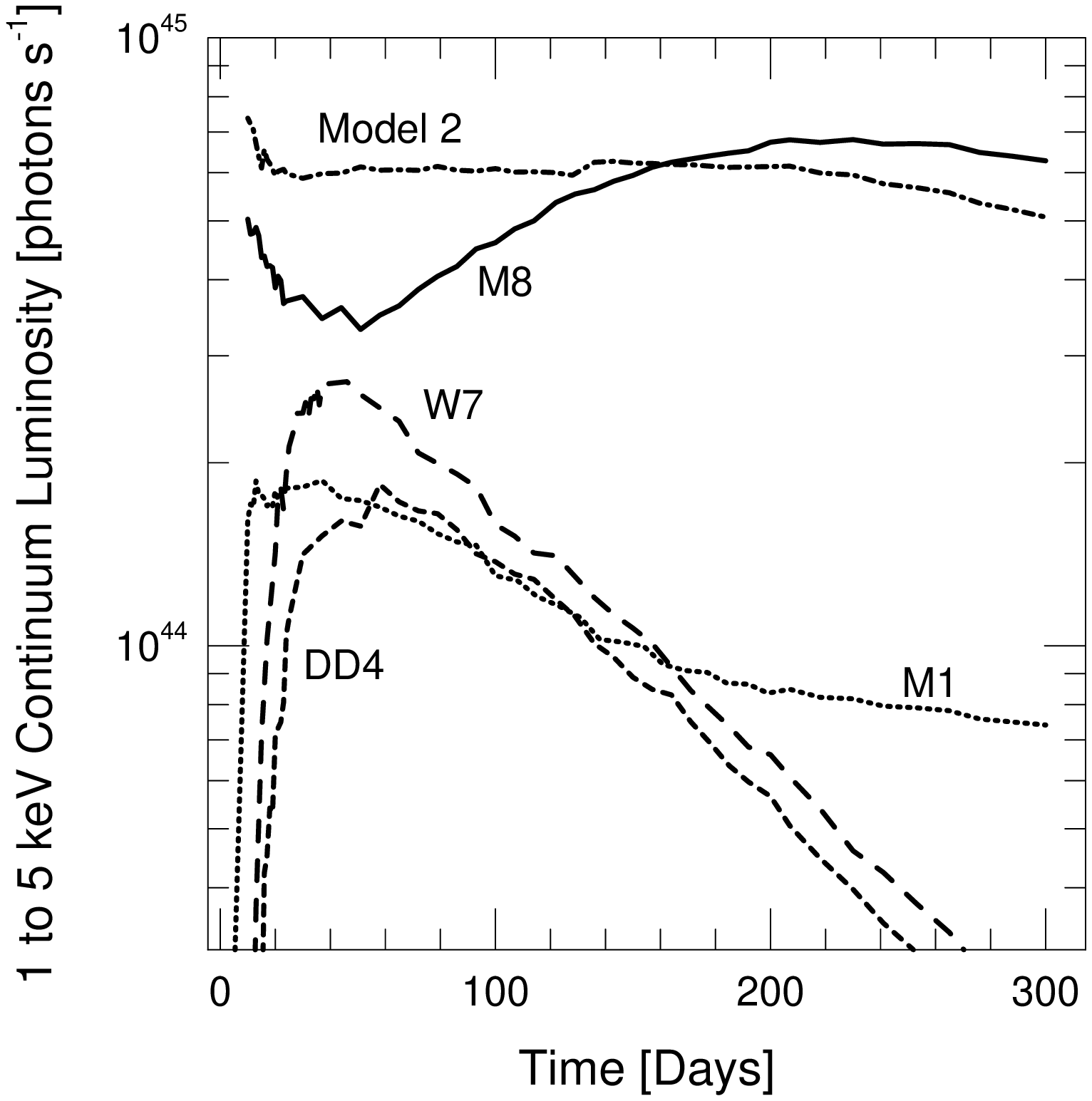}
\caption{Luminosity evolution in the 1 to 5 keV continuum.}
\label{redLC}
\end{figure}

Below, we discuss prospects for observing the X-ray emission from
SN~Ia with current, upcoming, and proposed X-ray observatories. In the
case of the Chandra Observatory, the sensitivity begins to fall off
dramatically beyond 5~keV and at 7~keV is quite small (effective
aperture $\sim 100\ \hbox{cm}^2$). This precludes direct observation
of iron peak K-shell emission from all but very nearby SNe~Ia.  We
consider the possibility of observing a SN~Ia at lower energy with
Chandra; Figure~\ref{redLC} shows the 1-5~keV luminosity evolution for
each of the models. Here we find that the faintest of the sub-\Mch
models, Model~M1, is indistinguishable from the two \Mch models during
the first 150 days. However, from 150 days onward, the light curve
flattens out and remains nearly constant for several hundred days
after explosion, whereas for the \Mch models it drops off, with an
e-folding time of $\sim130$ days. Off course, model M1 would be very
faint optically as well.

\begin{figure}[!t]
\plotone{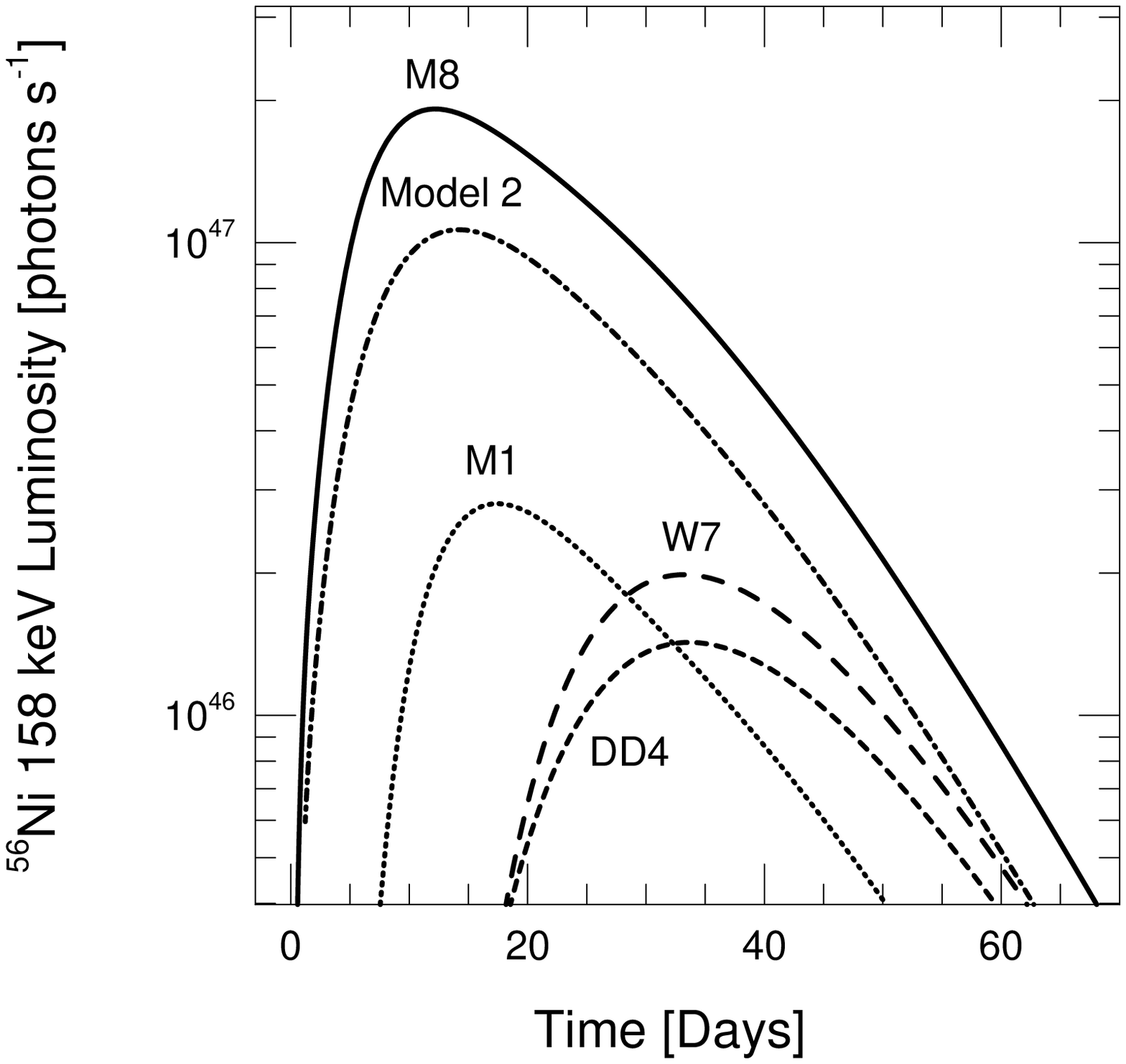}
\caption{The \nifsx 158 keV light curve for all models. The sub-\Mch
models are up to $\sim10$ times brighter, and peak substantially
earlier ($\sim10$~days) than the \Mch models ($\sim30$ days).}
\label{158keVLC}
\end{figure}

\begin{figure}[!t]
\plotone{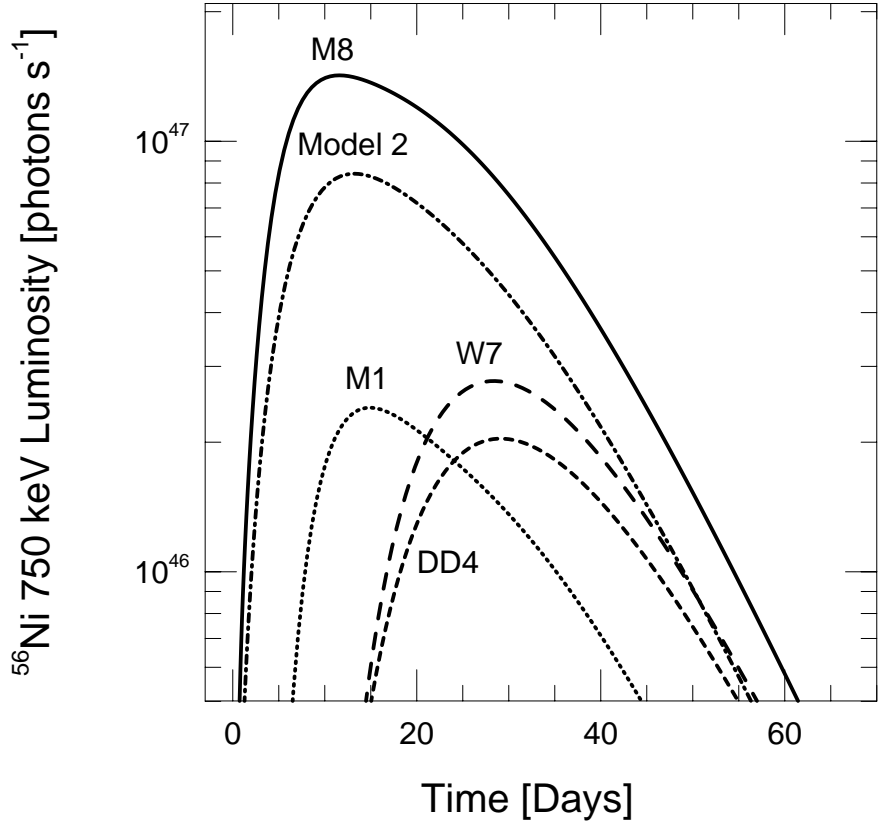}
\caption{Same as Figure \ref{158keVLC}, but for the \nifsx 750 keV line.}
\label{750keVLC}
\end{figure}

\begin{figure}[!t]
\plotone{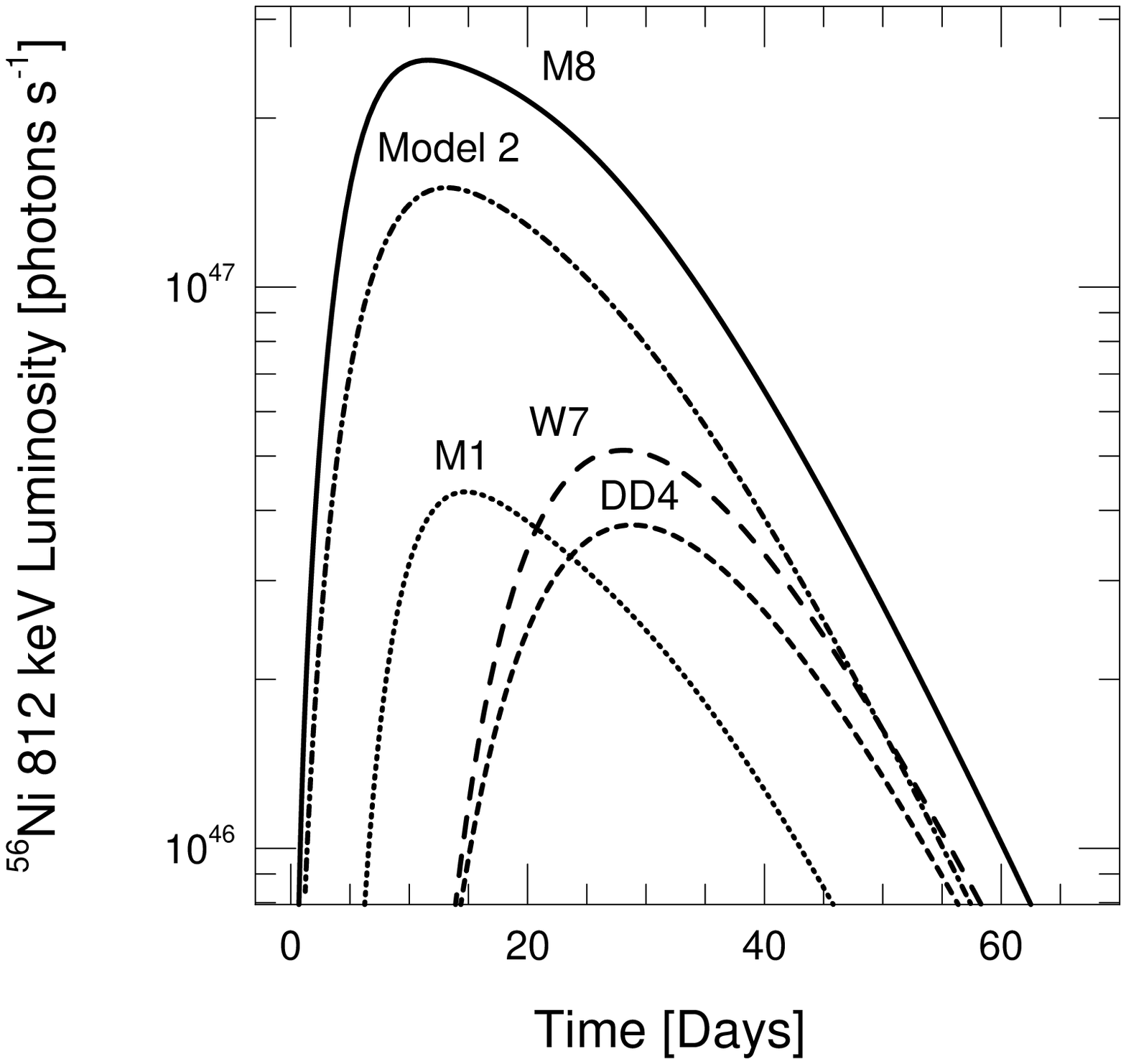}
\caption{Same as Figure \ref{158keVLC}, but for the \nifsx 812 keV line.}
\label{812keVLC}
\end{figure}

Most previous work on the $\gamma$-ray evolution of SNe~Ia focused on
the evolution of \Mch models \citep{GehrelsLM87,BurrowsT90}, and
especially on the \cofsx line evolution, since these have the greatest
chance of detection from an \Mch SNe~Ia. However, in the sub-\Mch
models the predicted \nifsx line fluxes are as much as 10 times
brighter than for models such as W7 and DD4, within a factor of two as
bright as their \cofsx lines.  The \nifsx 158 keV light curve for all
models is shown in Figure~\ref{158keVLC}. Once again, because the
Compton optical depth to core \nifsx remains high during the first few
\nifsx half-lives, the emergent flux in the sub-\Mch models is
dominated by decay of surface \nifsx.  Not only are the sub-\Mch
models brighter, but they also peak substantially earlier
($\sim10$~days) than the \Mch models ($\sim30$ days). Similar results
are obtained for all the other \nifsx decay
lines. Figures~\ref{750keVLC} and \ref{812keVLC} show the light curves
for the \nifsx 750~keV and 812~keV lines, respectively.

\citet{HoflichWK98} have also shown that \nifsx $\gamma$-ray lines peak
earlier and are brighter in sub-\Mch explosions. Indeed, it appears from
their Figure~8 that in the two sub-\Mch models they studied the \nifsx
lines reach maximum luminosity at time $t=0$. It is not clear how this can
be so. In the models studied in this paper, the exploded accretion layer
has sufficient Compton optical depth to delay the maximum to $\sim10$
days. \citet{HoflichWK98} also predict a substantially higher luminosity in
\nifsx lines. For instance, their Model~HeD6 is halfway between Model~2 and
Model~M8 in its initial mass but has nearly the same amount of
surface \nifsx (0.08~\Msun) as Model~2 (0.09~\Msun). For the \nifsx 812~keV
line they predict a maximum luminosity of $\sim8.5\times10^{47}\
\hbox{photons}\ {\rm s^{-1}}$. But for Model~M8, which has 0.17~\Msun of
\nifsx on the surface, we obtain only $\sim3\times10^{47}\ \hbox{photons}\
{\rm s^{-1}}$ at maximum, 10 days after explosion. We cannot explain this
discrepancy; to achieve their line fluxes, the surface layers must expand
with a much higher velocity than in Models 2, M1, or M8.  We note that our
line transport results have been verified by performing the calculations by
both Monte Carlo and deterministic methods, as described in the last
section; consistent results were obtained between the two methods in all
cases.

\begin{figure}[!t]
\plotone{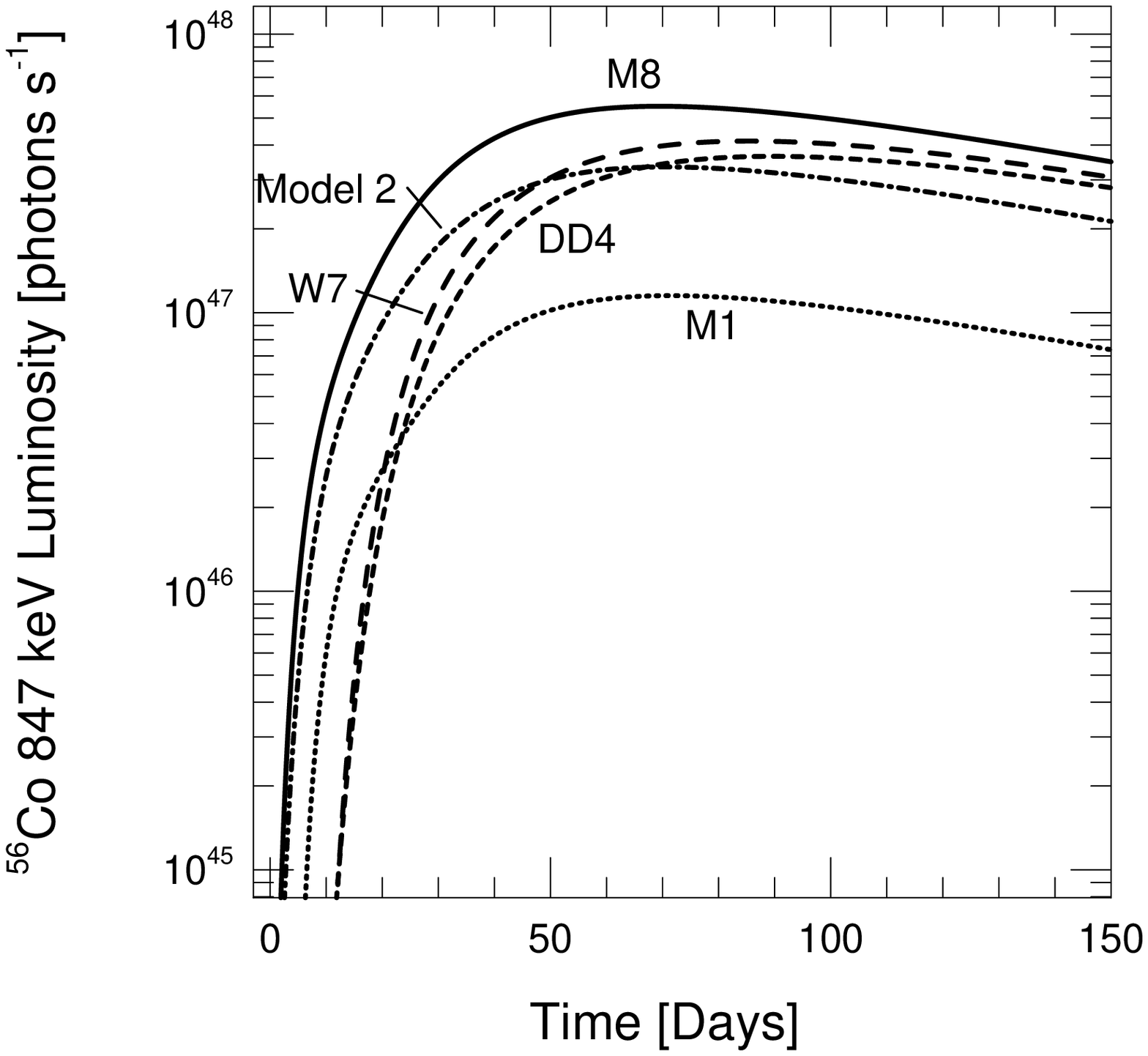}
\caption{The \cofsx 847 keV light curve for all models. Although the
sub-\Mch light curve rises more quickly than does the \Mch light
curves, the differences in maximum brightness depend only on the total
mass of \nifsx produced.}
\label{847keVLC}
\end{figure}

\begin{figure}[!t]
\plotone{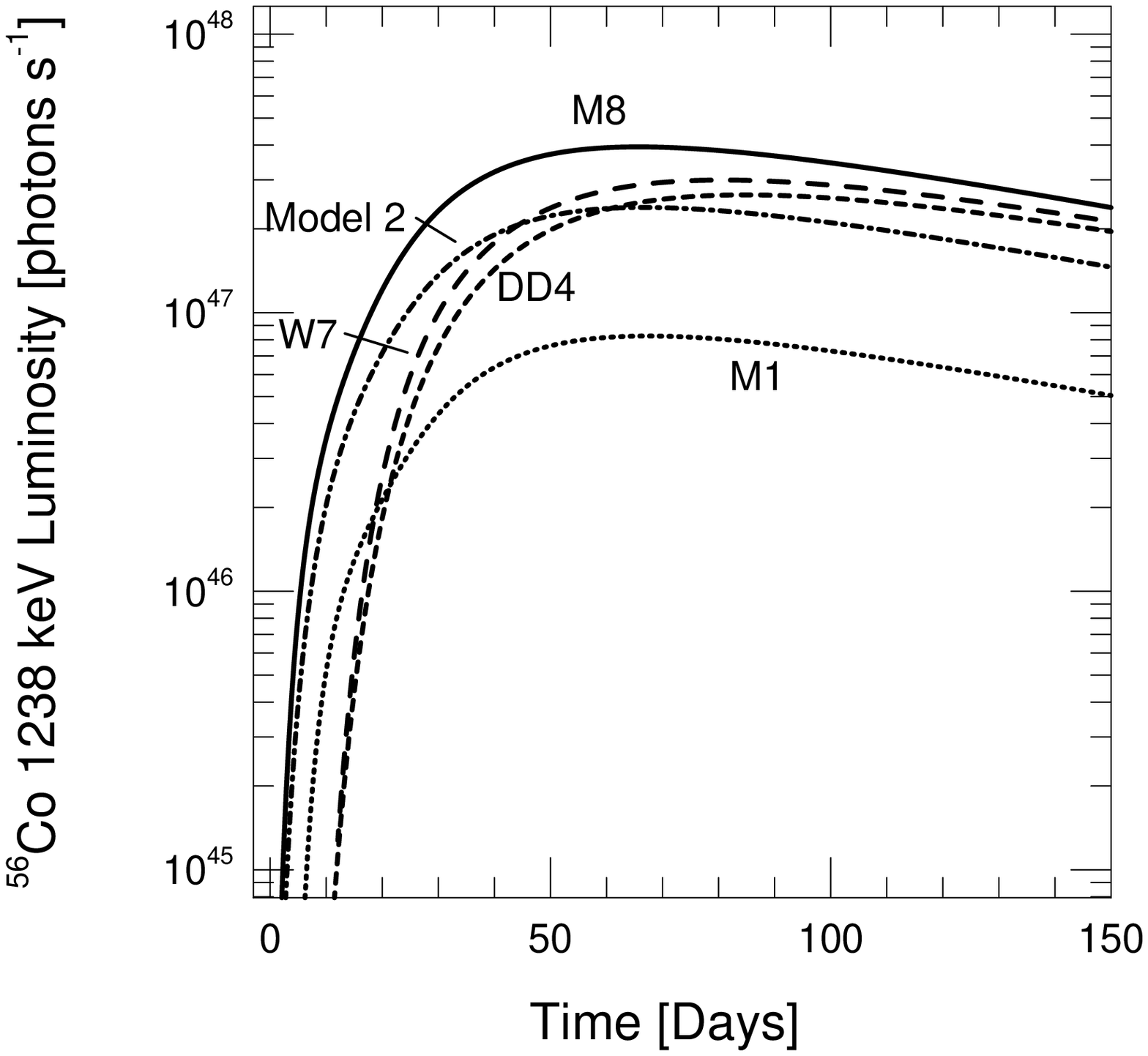}
\caption{Same as Figure \ref{847keVLC}, but for the \cofsx 1238 keV line.}
\label{1238keVLC}
\end{figure}

The situation is quite different for \cofsx lines. The light curves
for \cofsx 847~keV and 1238~keV are shown in Figures~\ref{847keVLC}
and \ref{1238keVLC}, respectively.  Since \cofsx is so much
longer-lived than \nifsx ($t_{1/2} = 77.1$~days versus 6.1 days), the
light curves for these lines reach maximum once the Compton optical
depth to the core is less than 1, which happens at around 50 days past
explosion.  The \Mch and sub-\Mch models are distinguishable by the
rise time, which is earlier in the sub-\Mch models, but this might be
difficult to detect, particularly for objects which are only
marginally within detection limits at maximum brightness. Differences
in the core \nifsx mass can mask the effect of any contribution from
surface \nifsx.

\section{Prospects for Detection}

Measurement of the 7~keV K-shell emission from Type~Ia supernovae provides
a direct and straightforward basis for discriminating between the two
current classes of progenitor models.  Detection of a large K-shell line
flux or bright bremsstrahlung continuum would indicate surface \nifsx,
which at present is predicted in significant quantities only from the
sub-\Mch model for Type~Ia supernovae. Additionally, the light curves of
\nifsx $\gamma$-ray decay lines from sub-\Mch models are distinctly
different than for \Mch models, providing an unambiguous indication of the
presence or absence of \nifsx at the surface.

Unfortunately, prospects are at best marginal for detecting X-ray
emission from a SN~Ia at a distance as great as Virgo using the
current generation of X-ray observatories. XMM, which was launched
10~December 1999, has the largest effective aperture at 7~keV of
current and near-term missions. We have used the XMM simulator
(SCISIM) to assess the likelihood of detection for the X-ray spectrum
of Model 2 at 100 days. Assuming a nominal distance of 15~Mpc, the
integrated 5-10~keV flux is $1.82\times10^{-7}\ \hbox{photons}\ {\rm
s}^{-1}\ \hbox{cm}^{-2}$.  Using both EPIC MOS CCDs and the PN CCD,
and assuming a half-power diameter of $15^{\prime\prime}$, 10 counts
would be detected in a $10^5$~second exposure. Although this may seem
like a small number, XMM's background count rate in the 5-10~keV range
is very low, ($\approx 1.7\times10^{-5}\
\mbox{counts}~\mbox{s}^{-1}$); the estimated background is 1.7 counts,
making this (formally) a $7\sigma$ detection. Although this will not
provide the the type of detailed kinematic information which, ideally,
one would like to have to confirm the origin of the counts, it is
nonetheless sufficient to distinguish with high confidence between the
\Mch and sub-\Mch models for Type~Ia supernovae.

The prospects for detecting a sub-\Mch SNe~Ia with the Chandra X-ray
Observatory are lower: at 7~keV the effective aperture of Chandra is
roughly $100\ \hbox{cm}^2$ for the ACIS-S imaging spectrometer; under
the previous assumptions this would give only 2 photons in a $10^5$
second exposure. The sensitivity is higher at lower energy, but the
supernova is also fainter. The total photon flux from Model~M8 at
15~Mpc is $3.4\times10^{-9}$ \photpersec in $1 \mbox{keV} < E < 2
\mbox{keV}$, and $1.5\times10^{-8}$ \photpersec in $2 \mbox{keV} < E <
5 \mbox{keV}$.  The effective aperture of the ASCA Imaging
Spectrometer in the $1 \mbox{keV} < E < 2 \mbox{keV}$ range is
approximately $600~\mbox{cm}^2$. Consequently, a $10^6$~second
observation would result in only 2~counts.  For the $2 \mbox{keV} < E
< 5 \mbox{keV}$ range, the effective aperture is approximately
$300~\mbox{cm}^2$ and the number of counts is approximately 4.5. Both
of these count rates would be lost in the noise, which is
substantially greater.


The proposed Constellation-X Observatory (Con-X: see
http://constellation.gsfc.nasa.gov) offers the best hope for high
resolution spectroscopy of K-shell emission from a sub-\Mch SNe~Ia in
Virgo. The proposal calls for Con-X's effective area to be
15,000~$\hbox{cm}^2$ at 1~keV, and 6,000~$\hbox{cm}^2$ at 6.4~keV.
Using the assumptions above, the number of counts received in
$10^5$ seconds would be 110 photons. Con-X might also be able to detect the
lower energy bremsstrahlung continuum. In Model~2 at 100 days, the
integrated 1-5~keV continuum flux at 15~Mpc would be $2\times10^{-8}\ {\rm
s}^{-1}$. Taking the mean aperture over this energy range to be
11,000~$\hbox{cm}^2$, the number of counts in a $10^{5}$ second exposure is
21 photons.

Immediate prospects for detecting \nifsx $\gamma$-line emission are
less promising. As an example, the peak 812~keV luminosity for
model~M8 is $\sim2\times10^{47}\ \hbox{photons}\ {\rm s}^{-1}$
(Figure~\ref{812keVLC}). At 15~Mpc this corresponds to a flux of
$7.4\times10^{-6}\ \hbox{photons}\ \hbox{cm}^2\ {\rm s}^{-1}$, which
is just {\it at} the sensitivity limit for INTEGRAL to detect narrow
lines in a $10^6$ second observation \citep{Winkler98}.  As
\citet{TimmesW97} have pointed out, flux levels such as produced by
the present set of sub-\Mch models could easily be seen with the
proposed ATHENA $\gamma$-ray Observatory \citep{Johnsonetal95}.

The view of \citet{HoflichWK98} for detecting \nifsx in sub-\Mch explosions
with INTEGRAL was slightly more optimistic than ours. However, as described
in the proceeding section, their predicted line luminosities are much
higher than ours, peaking almost immediately after explosion.

\section{$\gamma$-Line Profiles}

While neither current nor planned $\gamma$-ray spectroscopy missions
are sufficiently sensitive to obtain useful profile data from other than
a {\em very} nearby supernova, line profiles may some day serve as
another possible probe of the radial distribution of radioactivity.

\begin{figure}[!t]
\epsscale{1.0}
\plotone{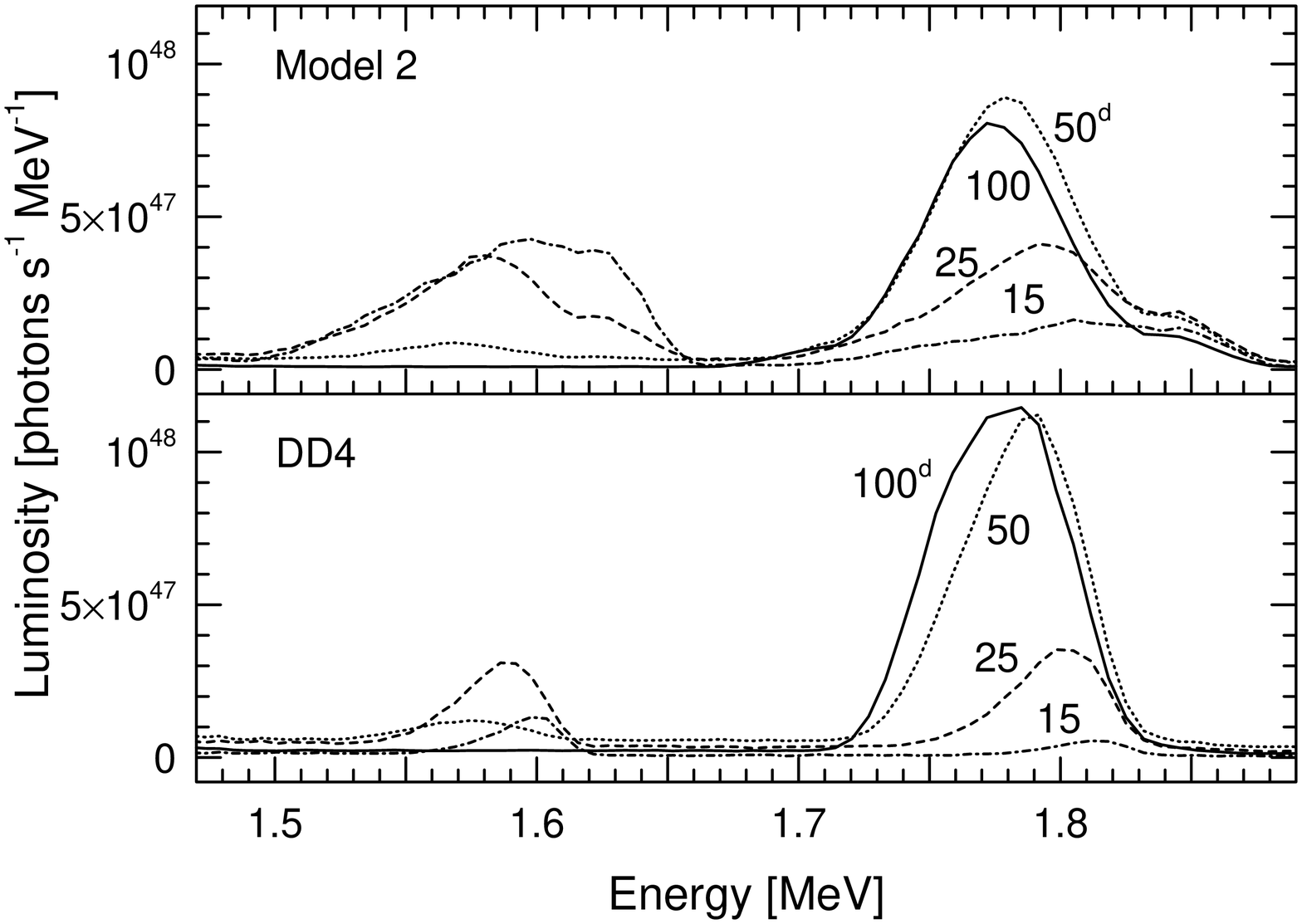}
\caption{Evolution of the \nifsx 1.562 MeV and \cofsx 1.772 MeV line profiles.
The top panel shows the evolution of these lines for the sub-\Mch model
hedtb11 at four different times; the bottom panel shows results from
the \Mch model DD4.}
\label{profile_fig}
\end{figure}

Figure (\ref{profile_fig}) shows the evolution of the \nifsx 1.562~MeV
and \cofsx 1.772~MeV line profiles. While these are weak lines, they
are representative, and their proximity in energy makes them a good
illustration of the difference in evolution of the \nifsx and \cofsx
emission. Because Compton scattering through any appreciable angle
removes photons from the profile altogether, the line shapes reflect
the velocity distribution of radioactive material at low optical depth
to the observer.

In the \Mch models, radioactive material is only found
deep within the supernova. Initially (15 days), the radiation which 
escapes is from material closest to the observer ($\gamma$-rays
emitted from material on the far side, away from the observer, see too
large an optical depth, and are absorbed). 
The peak intensity of the line is therefore blue-shifted by
$\sim 6000$\kms, and the line is very weak as a consequence of the
large optical depth. There is also a broad wing to the red resulting from
Compton scattering. As the ejecta expand and the column depth
decreases, a larger volume of the ejecta become visible and the red
side of the profile increases in intensity. By 100 days, the ejecta
are nearly transparent (optical depth to the center at 1 MeV is unity)
and the profile is that of an expanding optically thin sphere,
symmetric about the rest energy of the line. This is clearly seen in
the \cofsx profile, which reaches its maximum intensity at about this
time. The short half-life of \nifsx ensures that by the time the
profile has shifted back to the rest energy it has disappeared.

In the sub-\Mch models, the high velocity \nifsx layer on the outside
leads to \nifsx and \cofsx lines with rapid rise times (\cf\
Figures~\ref{158keVLC} and \ref{847keVLC}).  At the earliest times (15
days), the only optically-thin paths are those from the side
approaching the observer. The resulting line centers are blue-shifted
by 7000~\kms with wings extending up to 20,000~\kms to the blue. As
the ejecta expand (25 days), the optical depth through to surface
layers expanding on the opposite side of the supernovae becomes
smaller, and the red wings of the profiles increase in
intensity. While the core radioactivity is becoming visible, it
makes only a small contribution at this time.  By 50 days, however,
emission from the core dominates the profile, which is now virtually
identical to the \Mch model but for the low intensity but very broad
wings from the outer layer. In this model, the \nifsx lines are
proportionately much stronger at early times due to the low optical
depth to the surface layers; its short decay time ensures that the
\nifsx lines will always be observed to have higher energies than at
rest.

\section{Summary}

In this paper we have presented results of X- and $\gamma$-ray
transport calculations of \Mch and sub-\Mch explosion models for
Type~Ia supernovae. We have shown that the X-ray and $\gamma$-ray
spectral evolution of sub-\Mch models is distinctly different than
\Mch models. The presence of surface \nifsx in sub-\Mch supernovae
would make them extremely bright emitters of iron peak K-shell
emission, visible for several hundred days after explosion. K-shell
emission from a bright sub-\Mch located in the neighborhood of Virgo
would be just above the limit for detection by the XMM
Observatory. Detection by CHANDRA is unlikely for a SNe~Ia as distant
as Virgo.

Likewise, the \nifsx $\gamma$-ray light curves all display a
substantially different behavior in sub-\Mch and
\Mch models. Because of the presence of \nifsx on the surface,
the former are much brighter and peak much earlier than in \Mch
supernovae ($\sim 10$ versus $\sim 30$ days). The \nifsx $\gamma$-ray
line emission from a bright sub-\Mch explosion at 15~Mpc would be
just at the limit for detection by INTEGRAL.

Moderate resolution spectroscopy of $\gamma$-ray lines would allow
studying the evolution of line profiles. This would provide the most
detailed and readily interpreted data on the distribution of
radioactive elements and column density in SNe~Ia. Unfortunately,
there are at present no plans to construct an instrument with
sufficient aperture to observe any but the very closest (and rarest)
SNe~Ia in such detail.

\acknowledgments We greatly appreciate useful discussions with Dave
Arnett and Stan Woosley. We are particularly grateful to Patrick
Wojdowski for his invaluable assistance with the instrument
simulations, and to Lynn Kissel for his assistance with the
bremsstrahlung crossections.

This work was supported by the UC Lawrence Livermore National
Laboratory and the US Department of Energy (W-7405-ENG-48), and by the
National Science Foundation (CAREER grant AST9501634, PAP). PAP
gratefully acknowledges support from the Research Corporation though a
Cottrell Scholarship. TR gratefully acknowledges the support of a
graduate fellowship from the National Physical Sciences Consortium.

\end{document}